\documentclass[%
 reprint,
 amsmath,amssymb,
 aps,prapplied
]{revtex4-2}

\usepackage{graphicx}
\usepackage{dcolumn}
\usepackage{bm}

\usepackage{float}
\usepackage{braket}
\usepackage{complexity}
\usepackage[ruled,vlined]{algorithm2e}
\usepackage{algorithmic}
\usepackage[caption=false]{subfig}
\usepackage{xcolor}
\usepackage{grffile}

\begin{document}

\preprint{APS/123-QED}

\title{Bayesian phase estimation with adaptive grid refinement}
\thanks{A footnote to the article title}%

\author{Ramakrishna Tipireddy}
\email{Ramakrishna.Tipireddy@pnnl.gov}

 \altaffiliation[Also at ]{Physics Department, XYZ University.}
\affiliation{%
 Pacific Northwest National Laboratory, P.O. Box 999, MSIN K7-90, Richland, WA 99352
}%
\author{Nathan Wiebe}
\affiliation{%
 Pacific Northwest National Laboratory, P.O. Box 999, MSIN K7-90, Richland, WA 99352\\
 Department of Physics, University of Washington, Seattle WA 98195 
}%



\date{\today}

\begin{abstract}
We introduce a novel Bayesian phase estimation technique based on adaptive grid refinement method. This method automatically chooses the number particles needed for accurate phase estimation using grid refinement and cell merging strategies such that the total number of particles needed at each step is minimal. The proposed method provides a powerful alternative to traditional sampling based sequential Monte Carlo method which tend to fail in certain instances such as when the posterior distribution is bimodal. We also combine grid based and sampling based methods as hybrid particle filter where grid based method can be used to estimate a small but dominant set of parameters and Liu-West (LW) based SMC for the remaining set of parameters. Principal kurtosis analysis can be used to decide the choice of parameters for grid refinement method and  for sampling based methods. We provide numerical results comparing the performance of the proposed grid refinement method with Liu-West resampling based SMC. Numerical results suggest that the proposed method is quite promising for quantum phase estimation. It can be easily adapted to Hamiltonian learning which is a very useful technique for estimating unknown parameters of a Hamiltonian and for characterizing unknown quantum devices.

\end{abstract}

\maketitle


\section{\label{sec:level1}Introduction}
There has been a lot of progress on the development algorithms for quantum computing in recent years. Phase estimation is an important step in many quantum computing algorithms\cite{kitaev1995quantum,  berry2000optimal,  kitaev2002classical, o2009iterative, svore2013faster, paesani2017experimental, o2019quantum}.
In recent years there has been a push to move beyond simple phase estimation and instead learn more concrete models for states and processes.  This necessitates learning in higher-dimensional spaces and in turn bringing more sophisticated inference procedures to bear against the problem of phase estimation in the presence of device imperfections or more broadly techniques like quantum Hamiltonian learning \cite{granade2012robust, wang2017experimental, sergeevich2011characterization, wiebe2014quantum, wiebe2014hamiltonian}.

Bayesian methods have a long history of providing a logically consistent framework for solving such problems.  The two biggest benefits from using Bayesian methods are that adaptive inference protocols are straight forward to design in such frameworks and also the fact that the posterior distribution naturally quantifies the uncertainty in the inference.  The central challenge, however, is that direct application of Bayesian reasoning is exponentially expensive (owing to the curse of dimensionality).  The two most common ways that have been used in the field to circumvent these problems involve using Monte-Carlo approximations through sequential Monte-Carlo (otherwise known as particle filter methods) and also assumed density filtering which renders the problem efficient by making assumptions about the form of the prior and posterior distributions.

The approximations that underly these approaches will generically fail at some point in the inference procedure.  The most natural way of seeing this is from the fact that exact Bayesian inference is {\NP}-hard~\cite{cooper1990computational}. 
This means that unless $\BPP=\NP$, we cannot expect that efficient Bayesian inference is possible on probabilistic classical computers.  Regardless, there is a rich history of developing approximate learning methods that exploit structures in the data to be able to learn efficiently for many classes of problems but because of these complexity theoretic limitations care must be taken to ensure that the assumptions that underly each approach (such as Gaussianity or unimodality of the posterior) are met.  
Developing robust and cost effective methods that can deal with such problems therefore becomes critical if we intend to move beyond learning algorithms that are hand tuned by experts and towards automating such learning procedures.

In this paper we propose a new approach for phase estimation in the presence of noise and also in turn Hamiltonian learning that combines existing particle filter approaches with adaptive grid methods for approximating the posterior probability density.  We find numerically that these methods can be highly efficient and further do not suffer the same multi-modality issues that popular methods such as Liu-West resampling faces.

 \subsubsection{Bayesian Phase Estimation}
 Bayesian phase estimation is an approach to iterative phase estimation wherein the knowledge about the eigenphases of a unitary operation are represented on a classical computer via a prior distribution over the unknown phase.  In particular, let us assume that we have a unitary of the form $e^{-iH t}$ such that for an initial state $e^{-iHt} \ket{\psi} = e^{-i\omega t}$.  This holds without loss of generality for any unitary; however, for simplicity we assume here that $t\in\mathbb{R}$ rather than the discrete case where $t\in \mathbb{Z}$ (unless fractional queries are used~\cite{sheridan2009approximating,berry2012gate,gilyen2019quantum}).
 
 Bayesian phase estimation has a number of advantages.  First, it can easily be made adaptive since it manifestly tracks the current uncertainty and gives the user ways of estimating the most informative experiments to perform given the current knowledge.  Second, it provides a well motivated estimate of the uncertainty of the phase estimation procedure.  Finally, the framework can be easily extended to allow inference of other parameters that may be impacting the estimate of the eigenvalue $\omega$, such as $T_2$ times or over-rotations.  The central drawback of Bayesian inference is that the current state of knowledge is exponentially expensive to store and manipulate which necessitates the use of approximate forms of Bayesian inference.  
 
 In order to understand these tradeoffs it is necessary to briefly review the formalism of Bayesian inference.  The central object behind Bayesian inference is the prior distribution.  The prior distribution for the unknown phase $\omega$ is a probability density function $P(\omega)$ such that $\int_{\omega'}^{\omega'+\Delta} P(\omega) \mathrm{d}\omega$ gives the probability of the unknown phase being in the range $[\omega',\omega'+\Delta]$. 
 
 The next object in Bayesian inference is the likelihood function, which gives the probability that a given outcome would be observed given a set of hypothetical parameters.  For ideal Bayesian phase estimation the parameters that dictate an experiment are the evolution time $t$ and (optionally) a inversion phase $x$.  The data returned is a measurement outcome on a qubit, which is either zero or one.  The likelihood of measuring zero or one is
 \begin{align} \label{eq:likelihood1}
    P(0|\omega;t,x) &= \cos^2((\omega -x) t/2), \nonumber\\
    P(1|\omega;t,x) &= \sin^2((\omega -x) t/2).
\end{align}

Once a measurement outcome is observed, Bayesian inference allows the posterior distribution to be computed over the unknown phase given the experimental result.  The expression for the updated distribution is
\begin{equation}
    P(\omega|d;t,x) = \frac{P(\omega) P(d|\omega;t,x) }{\int P(\omega) P(d|\omega;t,x) \mathrm{d}\omega }
\end{equation}
In practice, the optimal values of $t$ and $x$ can be found by optimizing the Bayes risk of an experiment and in practice a good estimate of this optimal experiment can be quickly found using the particle guess heuristic (PGH)~\cite{wiebe2016efficient}.

This gives a rule for updating a distribution, but provides  little guidance about how to choose the initial distribution over the parameters.  In practice, it is common to choose the distribution to be uniform but if prior knowledge is provided about the unknown phase that can (and should) be reflected in the choice of distribution.

Bayesian inference can be extended beyond this idealized setting to accommodate models of noise in the system.  This is addressed by changing the likelihood function and, if necessary, the dimension of the prior distribution.  In particular, let us assume that the system is subject to decoherence and that the decoherence time $T_2$ is not known perfectly.  This can be addressed by switching to a two-dimensional prior distribution $P(\omega,T_2)$.  The likelihood function in this case can be written as
\begin{equation} 
    P(d|\omega,T_2;t,x) = e^{-t/T_2} P(d|\omega;t,x) + \frac{(1-e^{-t/T_2})}{2}.
\end{equation}
In this case, it is important to note that the prior distribution is two-parameter and as such requires quadratically more grid points to represent than the one-dimensional case.  In general if we have $D$ dimensions in the prior, then the resulting distribution would require a number of points that is greater by a power of $D$.  This is to say that exact Bayesian inference requires a number of points that grows exponentially with the number of parameters learnt and as such this curse of dimensionality can only be overcome through the use of approximate methods.

\section{\label{sec:smc}Approximate Bayesian inference}
Sequential Monte Carlo methods are often used for Bayesian inference and phase estimation~\cite{wiebe2016efficient}. The idea behind these methods is to approximate the probability density as a sum of Dirac-Delta functions.  In particular, let us assume that we have an initial prior probability density $P(\omega)$ we then wish to find a set of $\omega_j$ and $w_j$ that approximate $P(\omega)$ such that
\begin{equation}
    P(\omega) \approx \sum_{j=1}^{N_{\rm part}} w_j\delta(\omega -\omega_j).
\end{equation}
Here the notion of approximation that is meant is that there exists some length $L$ and $\delta$ such that 
\begin{equation}
    \max_{\omega'}\left|\int_{\omega' -L/2}^{\omega'+L/2} P(\omega) \mathrm{d} \omega - \sum_{j: \omega_j \in [\omega'-L/2,\omega'+L/2]} w_j\right| \le \delta\label{eq:approx}
\end{equation}
These delta functions are traditionally called particles.  If we take the $\omega_j$ to be sampled from the original probability distribution and assign uniform weights $w_j$ to each particle then $\delta \in O(1/\sqrt{N})$ and in this sense the quality of the Monte-Carlo approximation to the prior improves as the number of particles increases.

In this discrete representation Bayes' rule for updating the prior distribution based on the observed result is straight forward.  Let us assume that we observe outcome $d=\{0,1\}$, which are the two measurements that we could observe in an iterative phase estimation algorithm.  If we define the likelihood to be $P(d|\omega_j;t,x)$ then Bayes' rule can be applied to update our particle weights $w_j$ based on the measurement result via
\begin{equation}
    w_j \mapsto \frac{w_j P(d|\omega_j; t,x)}{\sum_j w_j P(d|\omega_j ;t,x)}.
\end{equation}

A key observation is that Bayesian inference does not cause the locations of the particles, i.e. the $\omega_j$, to move.  It only causes the weights to change.  This can prove problematic as such algorithms are repeated because the weights tend to decrease.  Resampling is a standard approach that can be used to address these problems.  The idea behind resampling is to draw a new ensemble of $\omega_j$ with $w_j = 1/N$ such that the new ensemble of particles models the resampled probability distribution well.

The method that has become common place for quantum applications is Liu-West resampling.  This scheme is given below.

\begin{algorithm}[H]
\label{alg:LW}
\SetAlgoLined
Input $(\omega_i, w_i), \quad i \in {1, \cdots, N}$ \\
Resampling parameter $a$ \\
Output: updated particles and weights $(\omega'_i, w'_i), \quad i \in {1, \cdots, N}$ \\
Liu-West(($\omega_i, w_i), a$) \\
 mean $\mu_{\omega} = \sum_i \omega_i w_i$ \\
 $h = \sqrt{1-a^2}$ \\
 covariance $\Sigma_{\omega} = h^2 \sum_i (\omega_i^2 w_i - \mu_{\omega}^2)$ \\
for $i = 1 \cdots N,$\\
\quad draw $j$ with probability $w_j$\\
\quad $\mu_{\omega_i} = a \omega_j + (1-a) \mu_{\omega} $ \\
\quad draw $\omega'_i$  from $\mathcal{N}(\mu_{\omega_i}, \Sigma_{\omega})$ \\
\quad $w'_i = 1/N$ \\
return ($(\omega'_i, w'_i)$)

 \caption{Liu-West resampling algorithm}
\end{algorithm}

The idea behind it is that the algorithm draws a new ensemble of particles such that the posterior mean and covariance are approximately preserved.  This choice is motivated, in part, by the fact that the posterior mean is the estimate that minimizes the expected square error in the estimated parameters.  Preserving this quantity and the covariance matrix means that the both this estimate and the uncertainty in this estimate are preserved in the resampling algorithm.

A major challenge however, is that while the posterior mean corresponds to the optimal unbiased estimator for the true parameters only if one is interested in the mean-square error in the inferred parameters.  In cases, such as phase estimation, where there can be degeneracies between positive and negative frequencies the mean square error in the frequency does not adequately reflect the estimate that minimizes the expected error in the resultant model's predictions.  This can cause the resampler to fail spectacularly.  Our aim is to improve this by using a new technique called adaptive grid refinement.  

\subsection{\label{sec:grid}Adaptive grid refinement}
In this section we introduce an adaptive grid refinement strategy for Bayesian phase estimation. The idea we employ borrows from the spirit of particle filtering.  Specifically, the idea that we use is that the essential information about the posterior distribution that needs to be maintained in the distribution is the probability density.  However, unlike resampling methods such as Liu-West resampling, we do not assume that the salient information about the posterior distribution is encoded in the low-order moments of the posterior.  Instead, our goal is to adaptively build a discrete mesh that supports the probability distribution.  This approach, unlike Liu-West resampling, suffers from the curse of dimensionality.  Our central innovation is to show that the two methods can be used simultaneously and thereby allowing parameters that are hard to estimate with particle filters to be learned using an adaptive mesh.

The aim of the mesh construction algorithm is to divide a region in parameter space into a set of boxes in $C_i \subset \mathbb{R}^{D}$ such that if $x_i$ is the centroid of the box $C_i$ with volume $V_i$ then $\int P(x) \mathrm{d}x \approx V_i P(x_i)$.  That is to say, we identify that the grid is in need of refinement when the midpoint rule for integration fails over any box in the mesh.  We then resample the distribution by dividing each dimension for the box into two sub-regions resulting in $2^D$ new boxes.  In general, more advanced quadrature methods could be deployed such as Richardson extrapolations or Runge-Kutta methods.  Such approaches have shown to have value for solutions to partial differential equations~\cite{huang2010adaptive} however here we focus our explorations on refining the precision of the mesh rather than the underlying quadrature formula.

\subsubsection{The midpoint rule}
The simplest possible case of using an adaptive mesh to perform approximate Bayesian inference involves using the midpoint rule in one dimension to form a grid that approximates the probability distribution.  One of the benefits of using the midpoint rule here, apart from its simplicity, is the fact that it comes with an explicit error bound.  We use this error bound to determine when the mesh needs to be subdivided.

Integration of $M = \int_a^b f(t) dt$ through midpoint rule with $n$ equal grid intervals is 
\begin{equation}
    M_n = \frac{b-a}{n} (f(m_1) + f(m_2) + \cdots + f(m_n)),
\end{equation}

where 
\begin{equation}
    m_k = \frac{t_k+t_{k+1}}{2} = a + \frac{2k-1}{2n}(b-a).
\end{equation}

 If $-K_2 \leq f'' \leq K_2$ on the interval [a,b], then the error $E_n(M)=|M_n-\int_a^b f(t) dt|$ is bounded by 
 
 \begin{equation} \label{eq:error_bound}
     E_n \leq \frac{K_2(b-a)^3}{24n^2}.
 \end{equation}
 If we therefore set $n=1$ then our aim when choosing the segments is to ensure that the average error per unit length of the box is constant.  That is to say
 \begin{equation}
     \frac{E_n}{(b-a)} \le \frac{K_2 (b-a)^2}{24}.\label{eq:errdens}
 \end{equation}
 Therefore, ignoring the irrelevant factor of $24$, an appropriate rule for refining the grid in such a way as to ensure that the total error in the probability distribution within each box is at most $\epsilon$ is to take for each segment of length $l_i$, the lengths such that $f''(x_i) l_i^2 = \epsilon$ for each interval $i$.

 In the slightly more complex case where we perform the analysis on a higher-dimensional mesh, the midpoint rule can be used straight forwardly in a  recursive fashion on the integral over each dimension. However since the cost of doing so grows rapidly for $D>1$, we focus below on the one-dimensional case for clarity.
 
 The idea that we present introduces an adaptive mesh into the problem.  The advantage of an adaptive mesh, fundamentally, is that the number of points needed to accurately estimate the probability density within a region can be smaller than that required by Monte-Carlo sampling.  Indeed, under reasonable continuity assumptions about the underlying distribution these differences can be exponential~\cite{atkinson2008introduction}.  However, these methods suffer from the curse of dimensionality.  Our approach will provide a way to achieve both advantages for Bayesian phase estimation without having to deal with the drawbacks of either approach.
 
In order to construct this grid adaptively we use two concepts.  The first is that of grid refinement.  The idea behind our algorithm is to use an elementary mesh and use the remainder estimate for the midpoint rule to decide whether the precision of the mesh within a region needs to be increased or not.  All dimensions that are not on the mesh are resampled using the Liu-West resampling algorithm described earlier.
 
 \subsubsection{Grid Refinement}
 The first procedure that we will describe is grid refinement.  The approach we take in this is to determine if there are any cells in the adaptive mesh that violate the bound on the error density that is tolerable according to~\eqref{eq:errdens}. In particular, we compute the error density for $i^{th}$ grid element as $e_i=f_i'' l_i^2,$ where $f_i''$ is the second derivative of the function $f_i=\omega_i w_i$ and $l_i$ is the length of the grid element. We use second order central difference method once to compute the first derivatives (gradient) $f'$ twice to compute the second derivative $f''.$ That is
 \begin{equation}\label{eq:f_grad}
     f_i' = \frac{f(\omega_{i+1})-f(\omega_{i-1})}{2l_i} + O(l_i^2)
 \end{equation}
 and
 \begin{equation}\label{eq:f_hess}
     f_i'' = \frac{f'(\omega_{i+1})-f'(\omega_{i-1})}{2l_i} + O(l_i^2).
 \end{equation}
We give pseudocode for this procedure below in Algorithm~\ref{alg:adapt_refine}.
 \begin{algorithm}[H]
 \label{alg:adapt_refine}
\SetAlgoLined
Input: $\{\omega_i, w_i, e_{th}\}$ \\
\quad  $f_i = \omega_i w_i,  \quad i \in \{1, \cdots, N\}$ \\
\quad  $e_i = f_i'' l_i^2$ \\
\quad for $j = 1, \cdots N$ \\
\quad \quad if $e_{j}>e_{th},$ \\ 
\quad \quad \quad compute segment end points $p_j$ and $p_{j+1}$ \\
\quad \quad \quad $\omega^r_{j_1} = \frac{\omega_j-p_j}{2}$ and $\omega^r_{j_2} = \frac{p_{j+1}-\omega_j}{2}$ \\
\quad \quad \quad Set $w^r_{j_1} = w^r_{j_2} = w_j/2$ \\
\quad \quad \quad Delete particle $\omega_j$ \\
\quad \quad \quad Insert particles $\omega^r_{j_1}, \omega^r_{j_2}$ into $\{\omega_i, w_i\}$ \\
\Return $\{\omega_i, w_i\}$ 
 \caption{Adaptive grid refinement algorithm}
\end{algorithm}
In the current approach each grid element meeting the refinement criterion is split only once. In general, we can employ a multi-pass version where the split is continued until none of the resulting grid elements satisfy the refinement criterion. 
 
 \subsubsection{Adaptive grid merge criterion}
 The following heuristic can be useful for managing the memory of our protocol.  As the learning procedure progresses, certain regions of parameter space will be assigned low posterior probability and thus no longer need the same level of precision prescribed by the adaptive grid protocol.  Here we address this issue by providing a heuristic for merging cells in the grid when the probability in them becomes too small.  We give explicit pseudo code below in Algorithm~\ref{alg:adaptive_merge} for one implementation of this merging heuristic. 
 
\begin{algorithm}[H]
 \label{alg:adaptive_merge}
\SetAlgoLined
Input: $\{\omega_i, w_i, w_{th}\}$ \\
\quad for each adjacent elements $\omega_{j}$ and $\omega_k$ \\
\quad \quad compute end points $p_j, p_{j+1}$ and $p_k, p_{k+1}$ for $\omega_j$ and $\omega_k$ \\
\quad \quad if $\max(w_j,w_k) < w_{th}$ \\
\quad \quad \quad $\omega^m_{jk} = (\max(p_{j+1},p_{k+1}) + \min(p_{j},p_{k}))/2$ \\
\quad \quad \quad $w^m_{jk} = w_j + w_k$ \\
\quad \quad \quad Delete $\omega_j, \omega_k$ \\
\quad \quad \quad Insert $\omega^m_{jk}$ into $\{\omega_i, w_i\}$ \\
\Return $\{\omega_i, w_i\}$
 \caption{Adaptive grid merge algorithm}
\end{algorithm}
 
 We develop adaptive grid refinement particle filter algorithm that incorporates adaptive grid refinement algorithm~\ref{alg:adapt_refine} and adaptive grid merge algorithm \ref{alg:adaptive_merge} for Bayesian phase estimation. We provide the pseudo code for adaptive grid refinement particle filter in Algorithm~\ref{alg:adaptivegrid_pf}.
 
 \begin{algorithm}[h]
\label{alg:adaptivegrid_pf}
\SetAlgoLined
Define prior grid $\omega_{grid} = \{\omega_1, \omega_2, \cdots, \omega_N\}$ \\
$k = 0, w^k_0 = 1/N$,\\
for $k = 1 \cdots N_k,$\\
\quad $\omega_{grid} = \omega_{grid}$ \\
\quad Set expt to experimental parameters \algorithmiccomment{i.e., expt = PGH($\{\omega_i, w_i\}, k$)} \\
\quad $d_k$ = simulate-expt($\omega$, expt) \\
\quad weights $\hat{w}^k_i = \hat{w}^{k-1}_{i} p(d_k|\omega_i^k)$\\
\quad Normalize 
\quad $w_k^i = \frac{\hat{w}^i_k}{\sum_{j=1}^N \hat{w}^j_k}$ \\
\quad ($\omega^r_i, w^r_i$) = Grid-Refine Algorithm~\ref{alg:adapt_refine} \\
\quad ($\omega^m_i, w^m_i$) = Grid-Merge algorithm \ref{alg:adaptive_merge} \\
\quad ($\omega_i, w_i$) = ($\omega^m_i, w^m_i$) \\
Estimate $\hat{\omega} = \sum_{i=1}^N \omega_k^i w^i_k$
 \caption{Adaptive grid refinement particle filter}
\end{algorithm}

\subsection{\label{sec:hybrid}Hybrid grid and sampling smc}
In this section we consider using a hybrid grid and a sampling based method to estimate the frequency $\omega$ and the dephasing parameter $T_2.$ We use grid based approach to estimate the frequency $\omega$ and Liu-West resampling based SMC for estimation the parameter $\theta = 1/T_2.$ The likelihood functions $P(0|\omega, T_2)$ and $P(1|\omega, T_2)$ are given by
\begin{align}
    P(0|\omega) &= e^{-t/T_2}\cos^2(\omega t/2) + \frac{1-e^{-t/T_2}}{2}, \nonumber\\
    P(1|\omega) &= 1-P(0|\omega).
\end{align}
Hybrid adaptive grid based and Liu-West resampling based algorithm is shown in algorithm \ref{alg:hybrid_grid_lw}. In this approach the prior on the parameter $\omega$ is defined on an one dimension grid $\omega \in [\omega_l, \omega_u]$. For each gird value $\omega_i$ a Liu-West resampling based sequential Monte Carlo filter is used to estimate the paramter $\theta_i.$ Then the particles corresponding to the one dimensional grid for $\omega$ and particles for $\theta_i$ are stacked together to form a set of two dimensional particles ${\omega_l, \theta_l}.$ Corresponding weights $w_l$ are obtained as tensor product of the weights $\{w_i\}$ and $\{w_{i,j}\}$ corresponding to the particles $\{\omega_i\}$ and $\{\theta_{i,j}\}$ respectively. 

\begin{algorithm}[H]
\label{alg:hybrid_grid_lw}
\SetAlgoLined
Define prior grid for $\omega$, $\omega_{grid} = \{\omega_1, \omega_2, \cdots, \omega_{N_1}\}$ \\
$k = 0, w^k_{\omega,i} = 1/N_1$,\\
for $k = 1 \cdots N_k,$\\
\quad Define uniform prior on $\theta \sim U[\theta_a, \theta_b]$ \\
\quad for $i=1 \cdots N_1$ \\
\quad \quad $\{\hat{\theta}_{i,j}, \hat{w}_{i,j}\}$ = SMC-LW($\{\theta_{i,j}, w_{i,j}\}, \omega_i$) \\
\quad Stack $\{\omega_i, w_i\}$ and $\{\hat{\theta}_{i,j}, \hat{w}_{i,j}\}$ as $(\{\omega_l,\theta_l\}, w_l)$ \\
\quad Update weights $\hat{w}^l_k = \hat{w}^l_{k-1} p(d_k|\omega^l_k, \theta^l_k)$\\
\quad Normalize 
$w^l_k = \frac{\hat{w}^l_k}{\sum_{j=1}^N \hat{w}^j_k}$ \\
\quad ($\omega^r_i, w^r_i$) = Grid-Refine Algorithm~\ref{alg:adapt_refine} \\
\quad ($\omega^m_i, w^m_i$) = Grid-Merge Algorithm \ref{alg:adaptive_merge} \\
\quad ($\omega_i, w_i$) = ($\omega^m_i, w^m_i$) \\
Estimate $\hat{\omega} = \sum_{i=1}^N \omega_k^i w^i_k$ and 
$\hat{\theta} = \sum_{i=1}^N \theta_k^i w^i_k$ \\
 \caption{Hybrid grid refinement-LW sampling algorithm}
\end{algorithm}

\subsection{Computational Complexity}
The computational complexity of this algorithm is largely dictated by the number of calls needed to the likelihood function during an update.  Each particle in SMC methods requires $O(1)$ calls to the likelihood function to perform an update~\cite{granade2012robust}.  Therefore, the number of such calls needed to the likelihood function is in $O(N_1N_k)$ where $N_1$ is the number of grid points that we consider and $N_k$ is the number of particles that we include in our filter at each point.  The error in the mean over the particle filter is in $O(1/\sqrt{N_k})$ in the worst case scenario where all the weight is only placed on one of the grid points.  The error in the mean over the grid is $O(K_2/N_1^2)$ from~\eqref{eq:error_bound}.  Assuming we begin from a uniform prior and experiments are performed with times $t_1,\ldots,t_M$ we then have that the posterior distribution for a uniform prior on the continuum is proportional to the likelihood function.  The second derivative of which with respect to the unknown parameter is, from the triangle inequality, at most $K_2 \in O((\sum_i |t_i|)^2)$.  Thus if we desire the error in the mean to be at most $\delta$ in the max-norm then it suffices to pick 
\begin{align}
    N_1 \in O\left(\frac{\sum_i |t_i|}{\delta} \right),~\qquad N_k \in O\left(\frac{1}{\delta^2} \right).
\end{align}
This implies that the cost of performing the protocol in terms of the number of likelihood evaluations is in $O(\sum_i |t_i| / \delta^3)$.  If the phase estimation procedure is Heisenberg limited then $\sum_i |t_i| \in O(1/\delta)$~\cite{daryanoosh2018experimental} which yields that the number of likelihood evaluations is in $O(1/\delta^4)$.

This worst case analysis shows that the algorithm is efficient, irrespective of the dimension of the model space orthogonal to the grid (if a constant value of $\delta$ suffices).  However, this estimate radically over-estimates the cost of most applications.  This is because most approaches to iterative phase estimation (the method of~\cite{svore2013faster} being a notable exception), use a mixture of short and long experiments that typically lead to a distribution that is supported only over a small fraction of the parameter space~\cite{wiebe2016efficient,granade2012robust,kimmel2015robust,dinani2019bayesian}.  This means that the worst case assumptions used above do not hold, and we anticipate empirically that the number of likelihood evaluations needed should be dominated by $N_k$ in such cases.  In particular, by choosing a grid that mimics the points evaluated using a high-order integration formula we can evaluate the integral over the grid within error $\delta$ using a polylogarithmic number of likelihood evaluations~\cite{hildebrand1987introduction}.  In such cases the complexity will be dominated by the SMC costs, which cause the costs to reduce to $\widetilde{O}(1/\delta^2)$.

\subsection{Principal Kurtosis Analysis} \label{sec:pka}
Given the cost of using a high-dimensional adaptive grid, a question remains about how one could use these ideas presented here for phase estimation more broadly for parameter estimation.  One approach that can be used to address this is a method known as principal kurtosis analysis~\cite{pena2010eigenvectors,meng2016principal}.  Kurtosis is a measure of the fourth moment of a distribution and measures the extent to which a distribution deviates from Gaussianity.  In particular, we can define a matrix that describes the Kurtosis of a random variable $x$ in a basis independent fashion via
\begin{equation}
    B=\mathbb{E}(((x-\mu)' \Sigma^{-1} (x-\mu))^2),
\end{equation}
where $\mu$ is the mean and $\Sigma$ is the covariance matrix of $x$.

The central idea behind principal Kurtosis analysis is to take a data set and find the directions that the data most strongly deviates most strongly from Gaussianity in by diagonalizing the matrix $B$ and selecting the components that have the largest eigenvalues.  This allows us when a resampling step would be invoked to choose the meshed axes of our posterior distribution to align with the directions of greatest kurtosis.  This allows us to adaptively choose the direction that we apply the mesh to during the resampling step and apply Liu-West resampling only to the directions with low excess Kurtosis.
\section{\label{sec:numerical}Numerical results}
In this section we present several numerical results to support adaptive grid refinement and hybrid grid and sampling based particle filters for quantum phase $\omega$ and dephasing parameter $T_2$ estimation. We show that with adaptive grid refinement particle filter, number of particles required can be controlled using a threshold $w_{th}$ for grid merging. We show the results for four different threshold values $10^{-5}, 10^{-4}, 10^{-3}$ and $2\times 10^{-3}.$ The threshold for grid refinement is fixed at $10^{-10}.$ In this section we show the numerical results for three different cases, namely i) quantum phase estimation in the absence of dephasing noise ii) quantum phase estimation in the presence of dephasing noise and iii) quantum phase and dephasing parameter estimation. 
\subsection{Quantum phase estimation without dephasing noise}
Here we consider the problem of quantum phase estimation in the absence of dephasing noise. The likelihood function for this problem are given by \eqref{eq:likelihood1}. For this problem we show quantum phase estimation results using the proposed adaptive grid refinement particle filter method and compare these results with standard Liu-West resampling based sequential Monte Carlo method. We consider two cases when a) $\omega \in [0, 1]$ and b) $\omega \in [-1,1].$ 
\begin{figure*}[t]
\centering
\subfloat[]{
\includegraphics[width=0.48\textwidth]{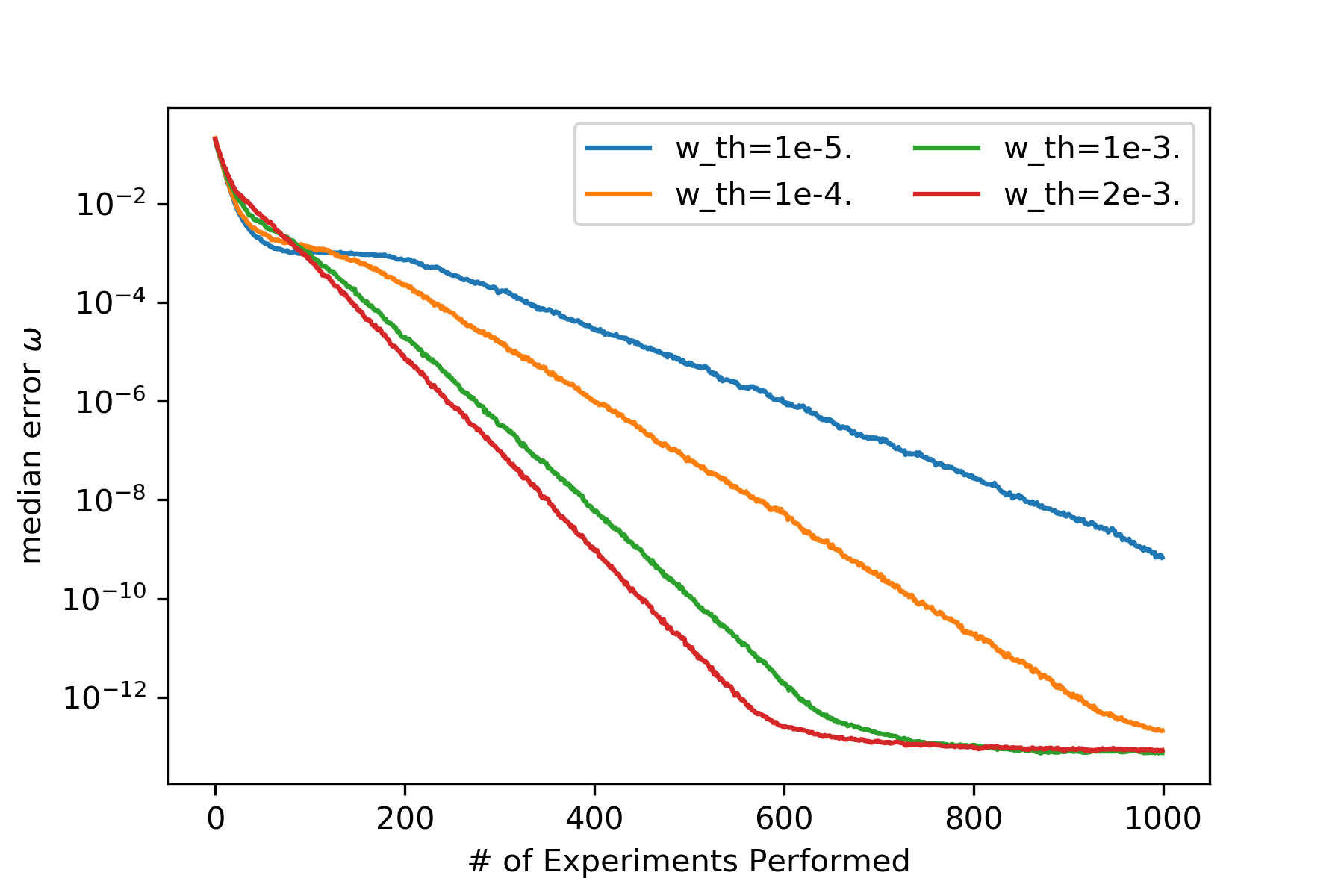}} 
\subfloat[]{
\includegraphics[width=0.48\textwidth]{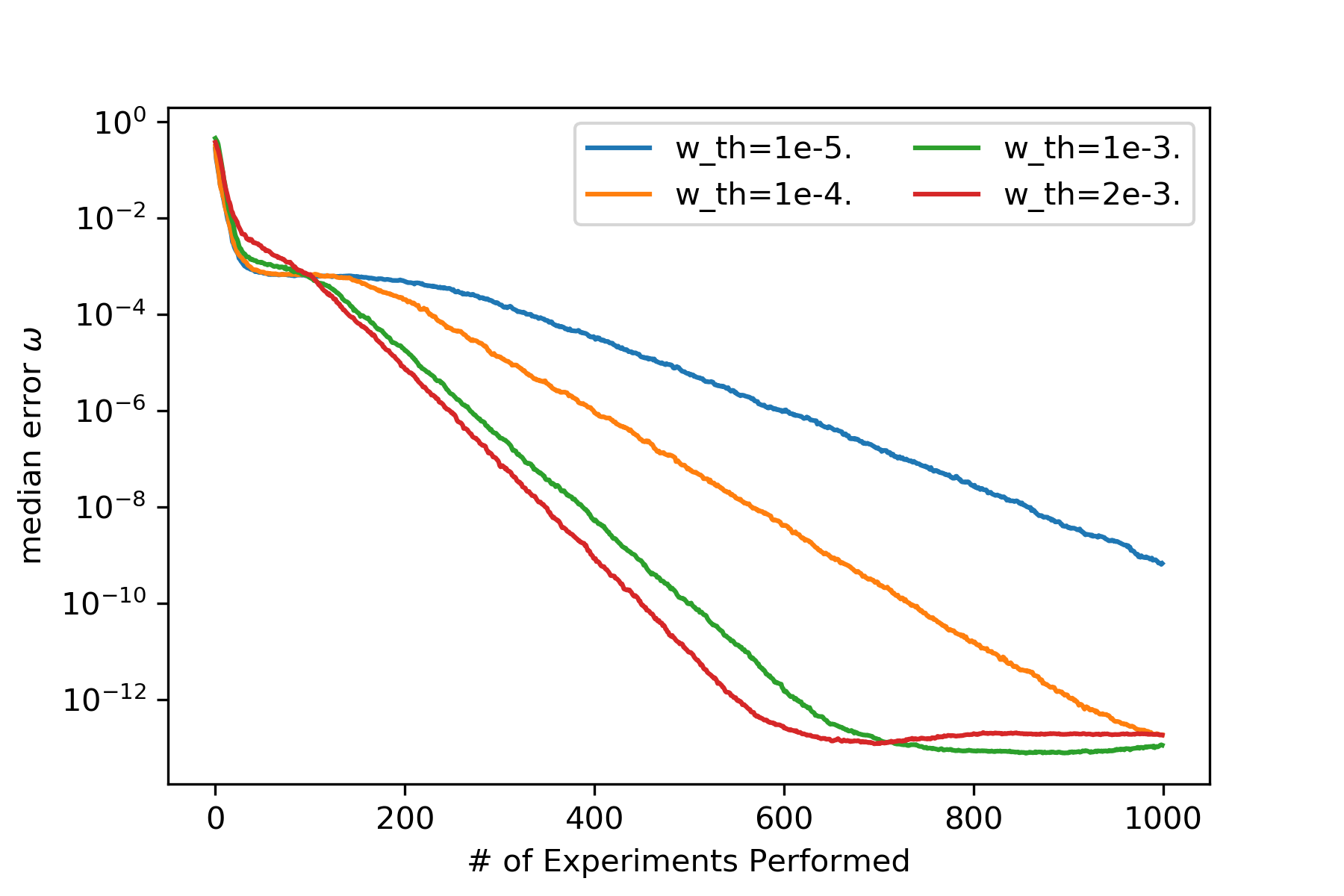}} \\
\subfloat[]{
\includegraphics[width=0.48\textwidth]{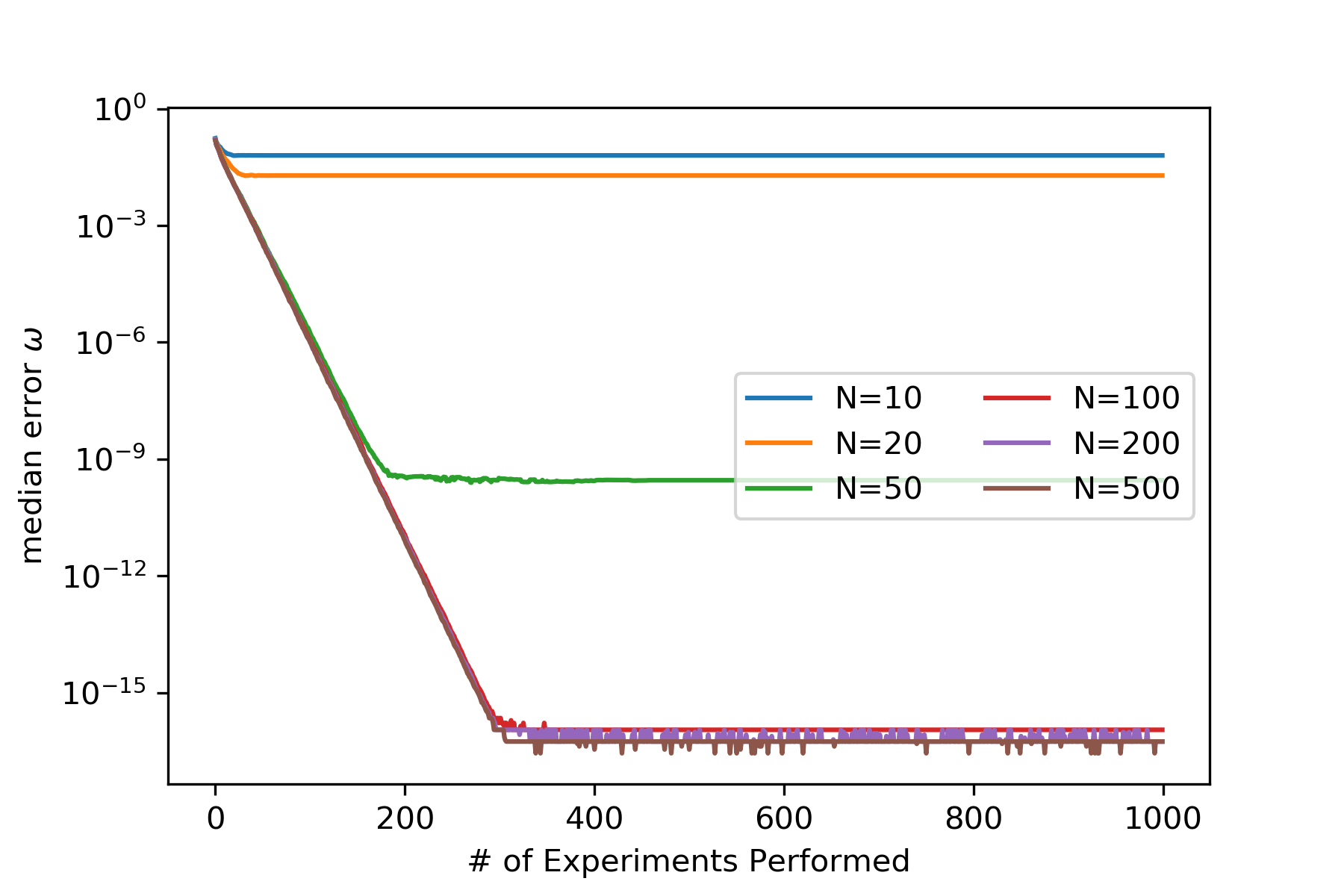}} 
\subfloat[]{
\includegraphics[width=0.48\textwidth]{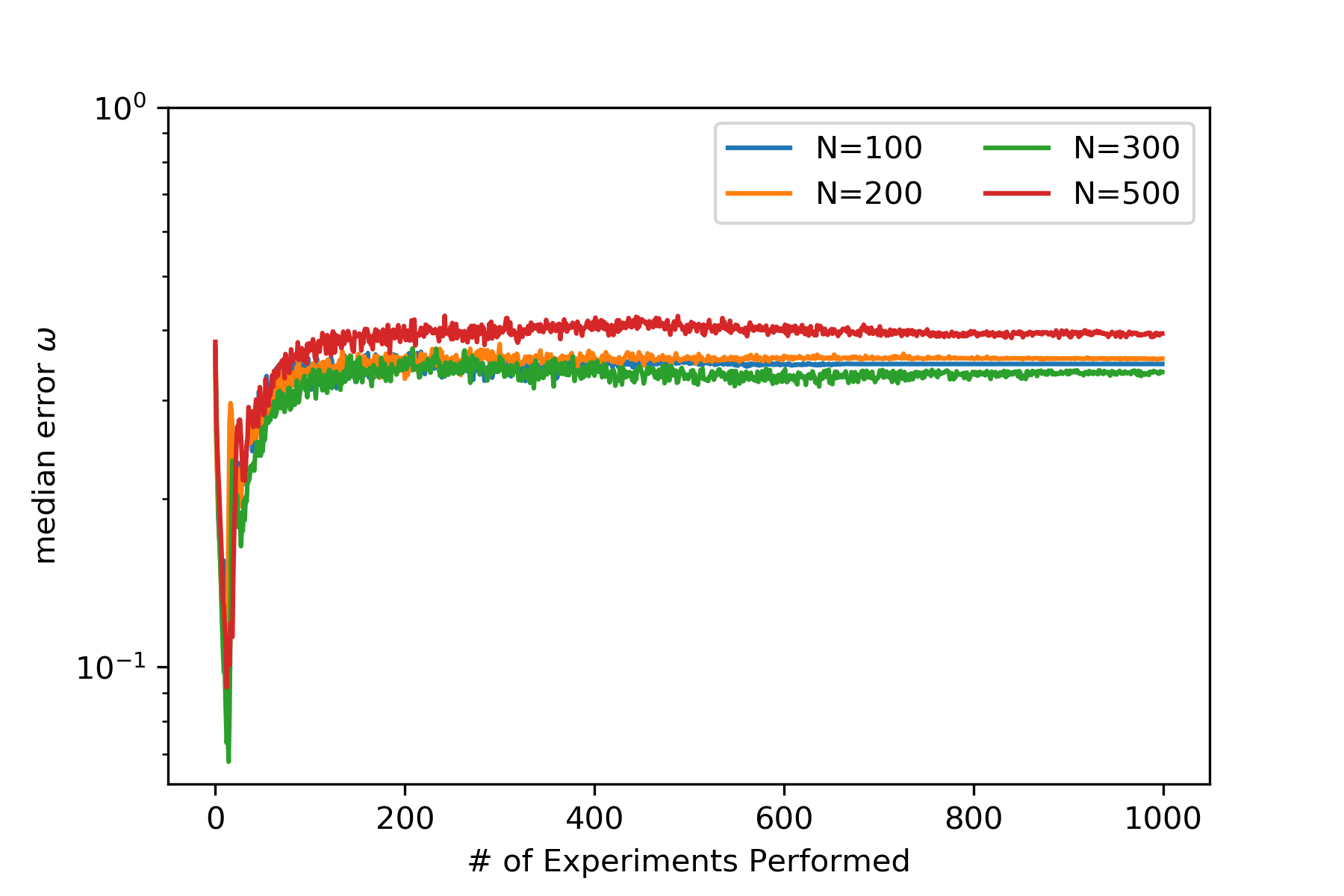}}
\caption{\label{with_1d_grid_median_error} Median  error for 1000 random a) $\omega \in [0,1]$, b) $\omega \in [-1,1]$ using adaptive grid refinement algorithm. Threshold for merging is $10^{-5}, 10^{-4}, 10^{-3}, 2\times 10^{-3}$. Median of MLE error using LW resampling algorithm for 1000 random c) $\omega \in [0,1]$, d) $\omega \in [-1,1]$ with number of particles $N=100, 200, 300, 500.$ }
\end{figure*}
For each case we perform simulations for $1000$ experiments (meaning that we perform a measurement, do a Bayesian update and adaptively choose the next experimental time $t_i$ based on the updated posterior distribution~\cite{granade2012robust}).  We further repeat this for $1000$ random reference values of $\omega.$ That is, for each random reference $\omega$, we perform $1000$ experiments to estimate its value. At the end we get $1000$ error estimates as a function of the number of experiments corresponding to $1000$ randomly chosen reference $\omega.$

We show the convergence of median error in the estimated phase $\omega$ with respect to the true reference $\omega.$ Due the grid refinement and merging algorithms in the proposed method, final number of particles needed depend on the threshold values used for refinement and merging. Error convergence also depends on the threshold values. Fig.\ref{with_1d_grid_median_error} shows the convergence of median error in posterior estimate of $\omega$ for different threshold values for merging. Fig.\ref{with_1d_grid_median_error}a shows the error for estimate of $\omega \in [0,1]$ and Fig.\ref{with_1d_grid_median_error}b shows the error for estimate of $\omega \in [-1,1].$ In both cases we can clearly see that the median error decreases as a function of the number of experiments and that the convergence is faster for larger values of the merging threshold value $w_{th}.$ To compare the proposed method with the standard Liu-West based sequential Monte Carlo method we show the numerical results for phase estimation in Figs \ref{with_1d_grid_median_error}c and \ref{with_1d_grid_median_error}d. Fig \ref{with_1d_grid_median_error}c shows the error for estimate of $\omega \in [0,1]$ and Fig \ref{with_1d_grid_median_error}d shows the error for estimate of $\omega \in [-1,1].$ As mentioned in adaptive grid refinement method the posterior estimate is used when $\omega \in [0,1]$ and maximum likelihood estimate is used when $\omega \in [-1,1],$ because the posterior distribution in the later case results in bimodal distribution for which posterior mean estimation is not accurate. In Fig \ref{with_1d_grid_median_error}c convergence of median error for different choice of number of particles ($N=10, 20, 50, 100, 200$ and $500$s). Although the median error for $N=10$ and $N=20$ doesn't converge very well, for $N\geq50$ the error converges to $<1e-9$ at around $200-300$ experiments. Fig \ref{with_1d_grid_median_error}d shows median error convergence results when $\omega \in [-1,1].$ Although the maximum likelihood estimate is used in both cases, Liu-West resampling based SMC (Fig \ref{with_1d_grid_median_error}d) performs poorly compared to the proposed adaptive grid refinement method (Fig.\ref{with_1d_grid_median_error}b).

 In all cases considered for adaptive grid refinement number of measurements needed to reach a particular level of uncertainty scales as $O(\log(1/\epsilon)$.  This means, for example, that even with $w_{\rm th} = 10^{-5}$ we will require roughly $1500$ experiments (bits measured) will be needed to hit the limits of numerical precision that SMC with Liu-West can achieve in roughly $200$ experiments.  While this approach requires nearly eight times as much data, we will see the processing time is comparable or better than that of Liu-West to hit that same threshold reliably for the case where $\omega \in [0,1]$.  In the case of $\omega \in [-1,1]$, however, adaptive grid refinement is clearly a superior strategy since we see Liu-West resampling causes the algorithm to fail.
 
 The scalings that we observe above are, in fact, nearly optimal.  This can be seen from the fact that the particle guess heuristic uses an evolution time per experiment on the order of $T_j \in O(1/\epsilon_j)$ where $\epsilon_j$ is the (circular) posterior standard deviation at update $j$~\cite{wiebe2014hamiltonian}.  Since the posterior variance shrinks exponentially with this the adaptive algorithm,  
 the total evolution time needed obeys $\sum_j T_j \in O( \sum_j 1/\epsilon_j) \subseteq \sum_j\exp(O(j)) \subseteq \exp(O(\log(1/\epsilon))) = O(1/\epsilon)$ 
 which saturates the Heisenberg limit (which coincides with the Bayesian Cram\`er-Rao bound in the absence of dephasing~\cite{berry2000optimal,granade2012robust})) up to a constant and is therefore nearly optimal. However, for the case of SMC using Liu-West resampling we only notice logarithmic scaling if the number of particles, $N$, is sufficiently large. In settings where we want to automate the learning process, the robustness of adaptive grid methods allows it to be applied without having to begin with an exhaustive hyperparameter search to find sufficient values for $N$ and the resampling threshold. 
\begin{figure}[h]
\includegraphics[width=0.48\textwidth]{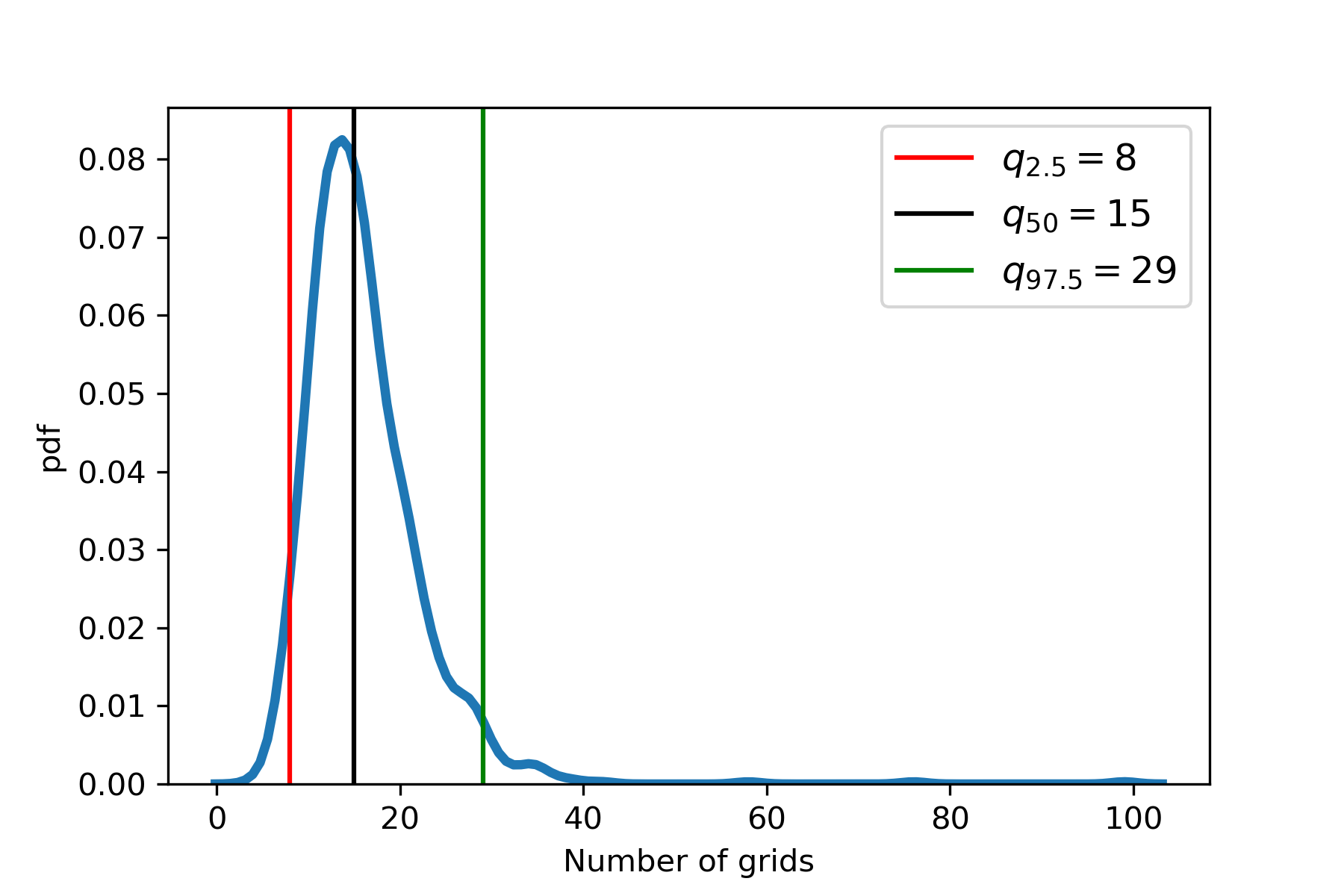}
\caption{\label{fig:pdf_plot} Probability density function of number of particles (grids) and 2.5 and 97.5 percentiles. Threshold for merging is $1e^{-5}$.}
\end{figure}
\begin{table}[h]
\begin{center}
\begin{tabular}{| p{1.5cm} | p{1.5cm} | p{1.5cm} | p{1.5cm} |}
\hline
 \multicolumn{4}{|c|}{Percentiles of number of grid points for $\omega \in [0,1]$} \\
\hline
 $w_{th}$ & $q_{2.5}$ & $q_{50}$ & $q_{97.5}$ \\ 
 \hline
 $1e^{-5}$ & 8 & 15 & 29 \\ 
 \hline
 $1e^{-4}$ & 8 & 19 & 136 \\ 
 \hline
 $1e^{-3}$ & 38  & 147 & 237 \\ 
\hline
 $2e^{-3}$ & 5 & 140 & 219 \\ 
\hline
\end{tabular} 
\caption{\label{tab:table-1} Percentile (2.5, 50 and 97.5) values of number of grid points required for estimating $\omega \in [0,1]$}
\end{center}
\end{table}

\begin{table}[h]
\begin{center}
\begin{tabular}{| p{1.5cm} | p{1.5cm} | p{1.5cm} | p{1.5cm} |}
\hline
 \multicolumn{4}{|c|}{Percentiles of number of grid points for $\omega \in [-1,1]$} \\
\hline
 $w_{th}$ & $q_{2.5}$ & $q_{50}$ & $q_{97.5}$ \\ 
 \hline
 $1e^{-5}$ & 8 & 15 & 29 \\ 
 \hline
 $1e^{-4}$ & 8 & 20 & 128 \\ 
 \hline
 $1e^{-3}$ & 20  & 141 & 238 \\ 
\hline
 $2e^{-3}$ & 11 & 141 & 219 \\ 
\hline
\end{tabular}
\caption{\label{tab:table-2} Percentile (2.5, 50 and 97.5) values of number of grid points required for estimating $\omega \in [-1,1]$}
\end{center}
\end{table}
The probability density function (pdf) and percentiles of number of particles needed for $1000$ random simulations is presented in Fig~\ref{fig:pdf_plot} for threshold value $w_{th}=10^{-5}.$ The median number of particles needed for this case is $15$ where as $2.5$ and $97.5$ percentile values are $8$ and $29$ respectively. Although the error convergence is slower for this case, the number of particles required is quite small compared to the Liu-West resampling case. For other threshold values of $w_{th}$ the median and the percentile values are presented in Tables \ref{tab:table-1} (for $\omega \in [0,1]$) and \ref{tab:table-2} (for $\omega \in [-1,1].$) When $\omega \in [-1,1]$ the posterior distribution is bimodal and hence the posterior estimate of $\omega$ does not provide accurate value. To alleviate this we use maximum likelihood estimate for this case. In Appendix~\ref{sec:appendix} Figs. \ref{with_1d_grid_refine_steps} (without dephasing noise) and \ref{with_dephase_1d_grid_refine_steps} (with dephasing noise) show the progress of grid refinement as we increase the number of experiments. We present detailed discussion in appendix~\ref{sec:appendix}. Although the grid refinement strategy is not efficient for higher dimensional problems due to exponential increase in the number of particles, it is convenient for the problems when the traditional sampling based methods fail. As described in section \ref{sec:hybrid}, we can combine both grid based and sampling based methods as hybrid methods to make use of advantages from both methods.

\subsection{Quantum phase estimation with dephasing noise}
\begin{figure}[t!]
\includegraphics[width=0.48\textwidth]{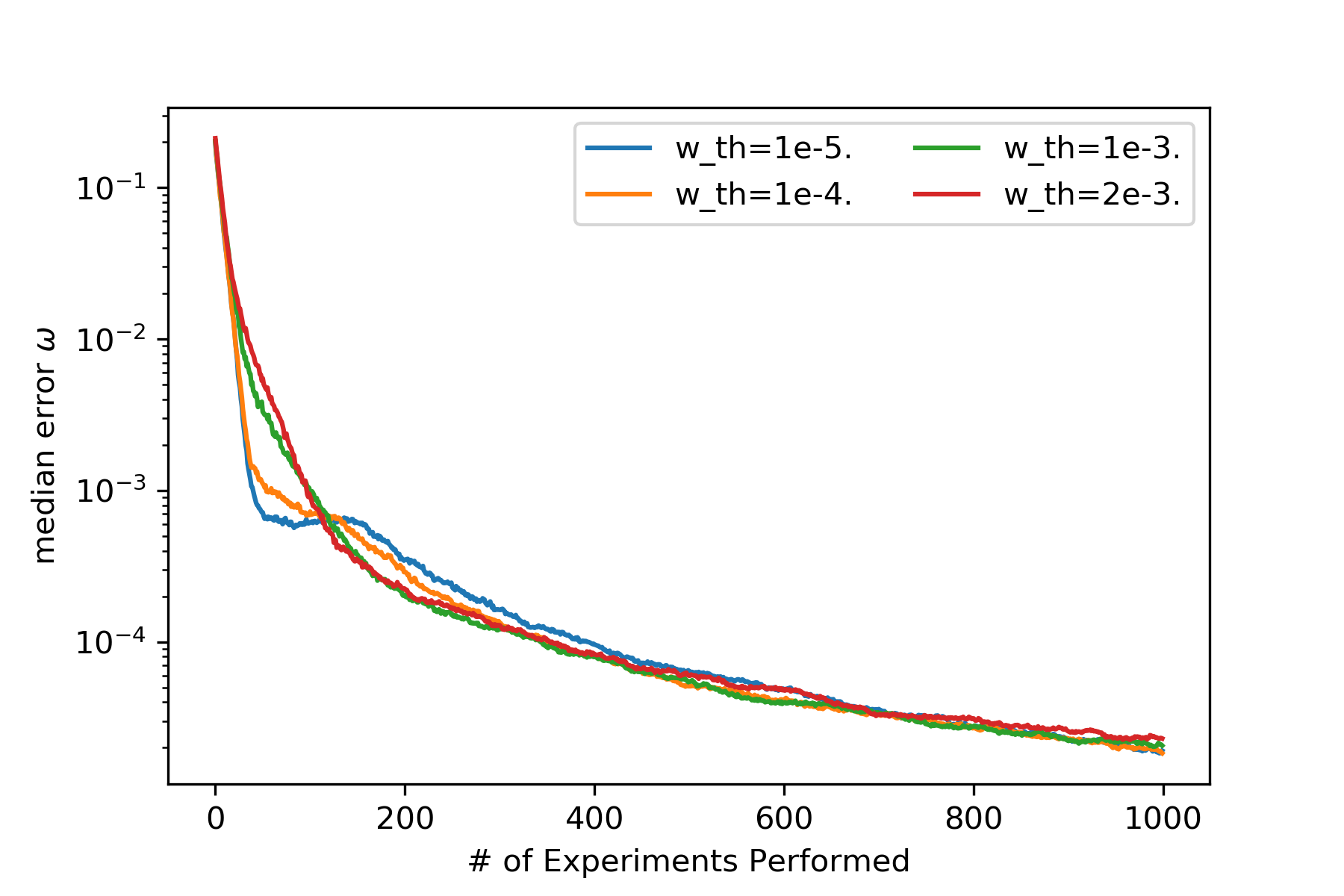}
\caption{\label{fig:dephase_omega_est} Adaptive grid refinement with dephasing noise. Median  error for 1000 random $\omega$, $T_2=50\pi$. Threshold for merging is $10^{-5}, 10^{-4}, 10^{-3}, 2\times 10^{-3}$.}
\end{figure}

As shown in the previous section the quantum phase estimation in the absence of dephasing noise is easier and high levels of accuracy ($10^{-12}$). However the quantum systems as are not perfect and quantum experiments are not accurate due to various disturbances such as dephasing noise. In this section we show the numerical results for quantum phase estimation in the presence of dephasing noise using adaptive grid refinement method. Because of the dehasing noise the information in the signal (output of the quantum experiment) is corrupted and hence the accuracy of the phase estimation is deteriorated. Fig \ref{fig:dephase_omega_est} the median error in the phase estimation as a function of the number of experiments for $1000$ random choices of true reference $\omega$ when the depahsing parameter is $T_2=50 \pi.$ As in the previous cases, the numerical experiments are performed for different choices of threshold $w_{th} = 10^{-5}, 10^{-4}, 10^{-3}, 2\times 10^{-3}$ used in the merging algorithm. As expected the accuracy ($\sim 10^{-5}$)of the algorithm in the presence of the dephasing noise is not as good as that we saw in the absence of the dephasing noise.  
\subsection{Quantum phase and dephasing parameter estimation using hybrid algorithm}
\begin{figure*}[t!]
\subfloat[]{
\includegraphics[width=0.48\textwidth]{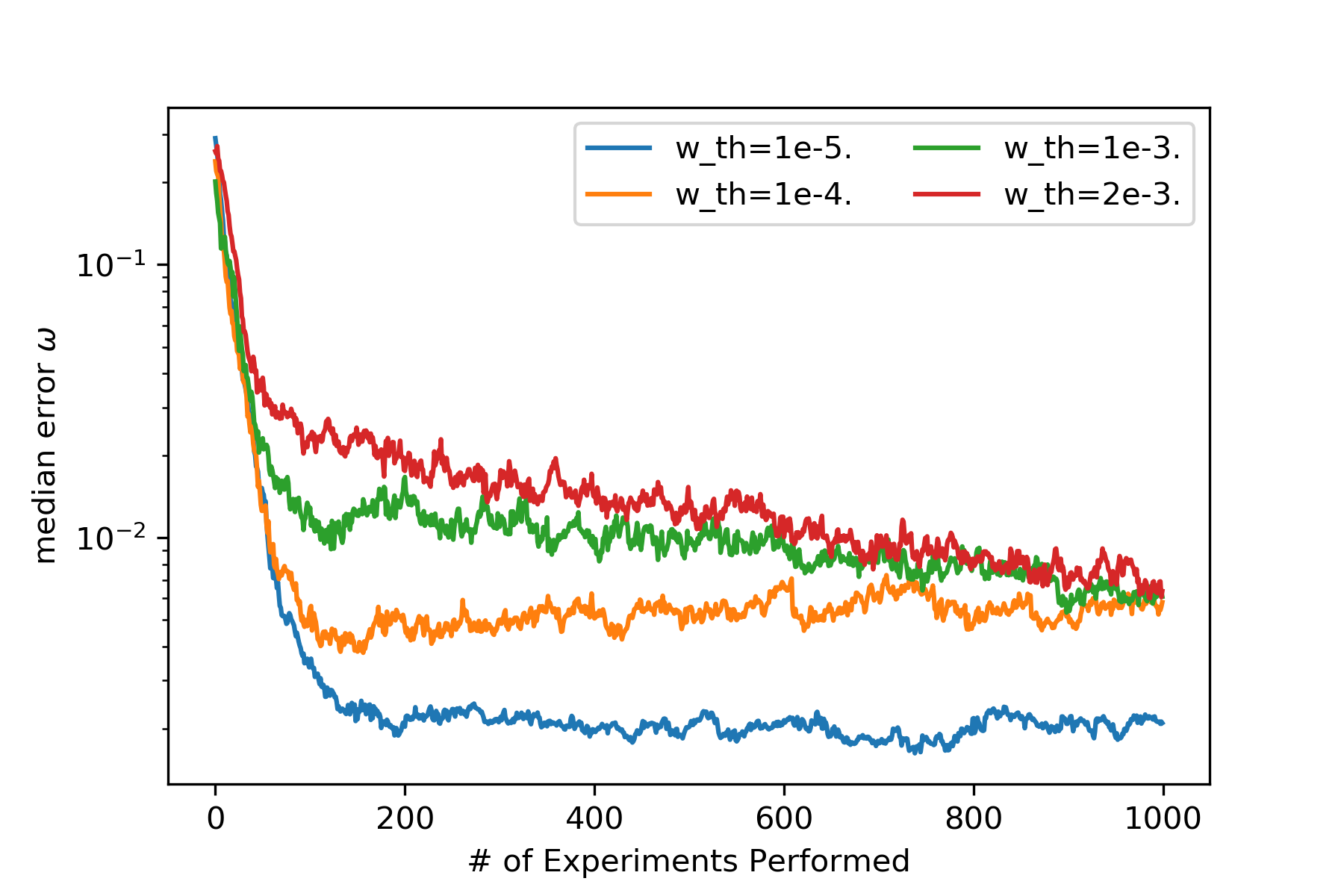}}
\subfloat[]{
\includegraphics[width=0.48\textwidth]{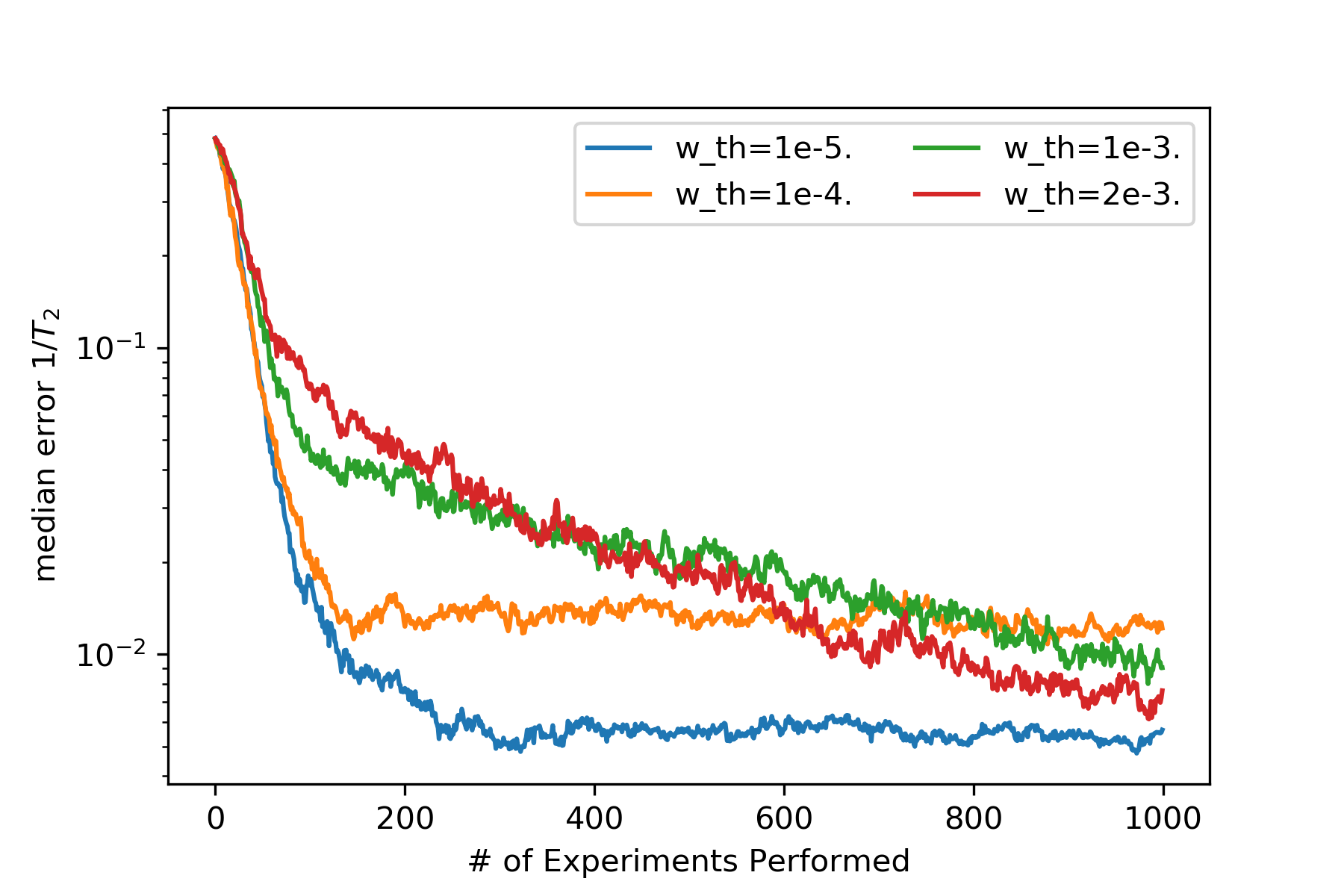}}
\caption{\label{fig:hybrid_omega_T2}Hybrid adaptive grid refinement and LW resampling method for a) phase ($\omega$) estimation and b) dephasing parameter ($T_2$). Median error in $\omega$ estimate for 100 random $\omega$ and fixed true $T_2=50\pi$. Threshold for merging is $10^{-5}, 10^{-4}, 10^{-3}, 2\times 10^{-3}$.}
\end{figure*}
In this section, we provide the numerical results using the hybrid algorithm described in section \ref{sec:hybrid} for estimating the quantum phase $\omega$ and the dephasing parameter $\theta=1/T_2.$ In this hybrid approach, we combine grid based and sampling based approaches such that the adaptive grid refinement is used to estimate the quantum phase $\omega$ and Liu-West resampling based SMC is used to estimate the dephasing parameter $\theta.$ In this method as shown in algorithm \ref{alg:hybrid_grid_lw}, a one dimensional grid in $\omega_l,\omega_u$ is used as a prior for $\omega$ and for each realization of $\omega_i$ as LW SMC is performed on the parameter $\theta_i~U[0,1].$ Then all the realizations $\{\omega_i\}$ and corresponding $\{\theta{i,j}\}$ are stacked together to form a set of particles in two dimensions. The Bayesian update is performed on this two dimensional particles. Although we arbitrarily chose grid refinement method for $\omega$ and Liu-West resampling for $\theta,$ in a multidiemnsioal setting, one can use principal Kurtosis analysis method described in section \ref{sec:pka} to to adaptively choose the direction that we apply the mesh to during the resampling step and apply Liu-West resampling to the directions with low excess Kurtosis. As shown in prior cases we perform numerical experiments for different choices of threshold $w_{th} = 10^{-5}, 10^{-4}, 10^{-3}, 2\times 10^{-3}$ for merging. Fig \ref{fig:hybrid_omega_T2}a shows convergence of median error in $\omega$ for $100$ random choices of true reference $\omega$ and for a given dephasing value of $T_2=50\pi$ obtained using hybrid algorithm. Fig \ref{fig:hybrid_omega_T2}b shows the median error in the parameter $\theta=1/T_2$ obtained using the same algorithms. Although the accuracy of the quantum phase estimate $\omega$ and the dephasing parameter $T_2$ are not as good as those we saw in the previous cases, the proposed approach provides as a powerful alternative when traditional sampling based methods fail one or more dimensions. This approach brings the advantages of both grid based and sampling based methods. We also observe that the effect of $w_th$ in Figs. \ref{fig:hybrid_omega_T2}a and \ref{fig:hybrid_omega_T2}b on the convergence rate is reversed with respect to the that in Fig. \ref{with_1d_grid_median_error}a. We believe that each problem is diffident need to optimize the hyper parameters (e.g. $w_th$, $e_th$) for each case. We will investigate this in the future.
\section{\label{sec:conclusions}Conclusions}
In this paper we present a novel Bayesian phase estimation method based on adaptive grid refinement method. We also combine grid based and sampling based methods as hybrid methods to simultaneously estimate the quantum phase and other parameters in the quantum Hamiltonian. We present numerical results quantum phase estimation for three different cases. In the first case we consider the phase estimation when the dephasing noise is absent. For this case we compare error convergence on the phase estimation using adaptive grid method and Liu-West resampling SMC method. We show that both adaptive grid and LW methods perform well when $\omega \in [0,1].$ However LW method performs poorly when $\omega \in [-1,1].$ We also provide numerical results for phase estimation using adaptive grid refinement when the dephasing noise is present. Finally we use hybrid algorithm to simultaneously estimate the quantum phase and the dephasing parameters which allows us to combine the benefits of both SMC and grid-based methods. 

There are many avenues for future work that are posed by this.  Although the curse of dimensionality will invariably limit the applicability of adaptive meshing schemes to low dimensional problems, the unprecedented accuracy that adpative meshing can provide can compensate for this.  Nonetheless, while the generalization of these results to adaptive grids in greater than one-dimension is conceptually straight forward, it is likely that with optimized implementations of meshing techniques used in finite element analysis and beyond will be able to make the technique computationally tractable in even three-dimensions.  Further, subsequent work remains in seeing the impact that rotating the grid to handle the dimensions of greatest Kurtosis in order to stabilize the Liu-West particle filter in the presence of multi-modality.  Such explorations are again conceptually straight forward but will require extensive numerical studies to understand the regimes where our techniques can provide an advantage.  It is our hope that by providing methods that can adapt the representation of the probability distribution we may be able to make phase estimation, and Hamiltonian learning more generally, robust and fully automatable. 

\begin{acknowledgments}
The material presented here is based upon work supported by the Pacific Northwest National Laboratory (PNNL) ``Quantum Algorithms, Software, and Architectures (QUASAR)". We also would like to thank Dr. Sriram Krishnamoorthy from PNNL for his valuable feedback and suggestions that have significantly improved the paper. PNNL is operated by Battelle for the DOE under Contract DE-AC05-76RL01830. 
\end{acknowledgments}

\bibliography{adaptivegridpe}


\appendix

\section{Additional Numerics} \label{sec:appendix}
\begin{figure*}[t!]
\centering
\subfloat[]{
\includegraphics[width=0.3\linewidth]{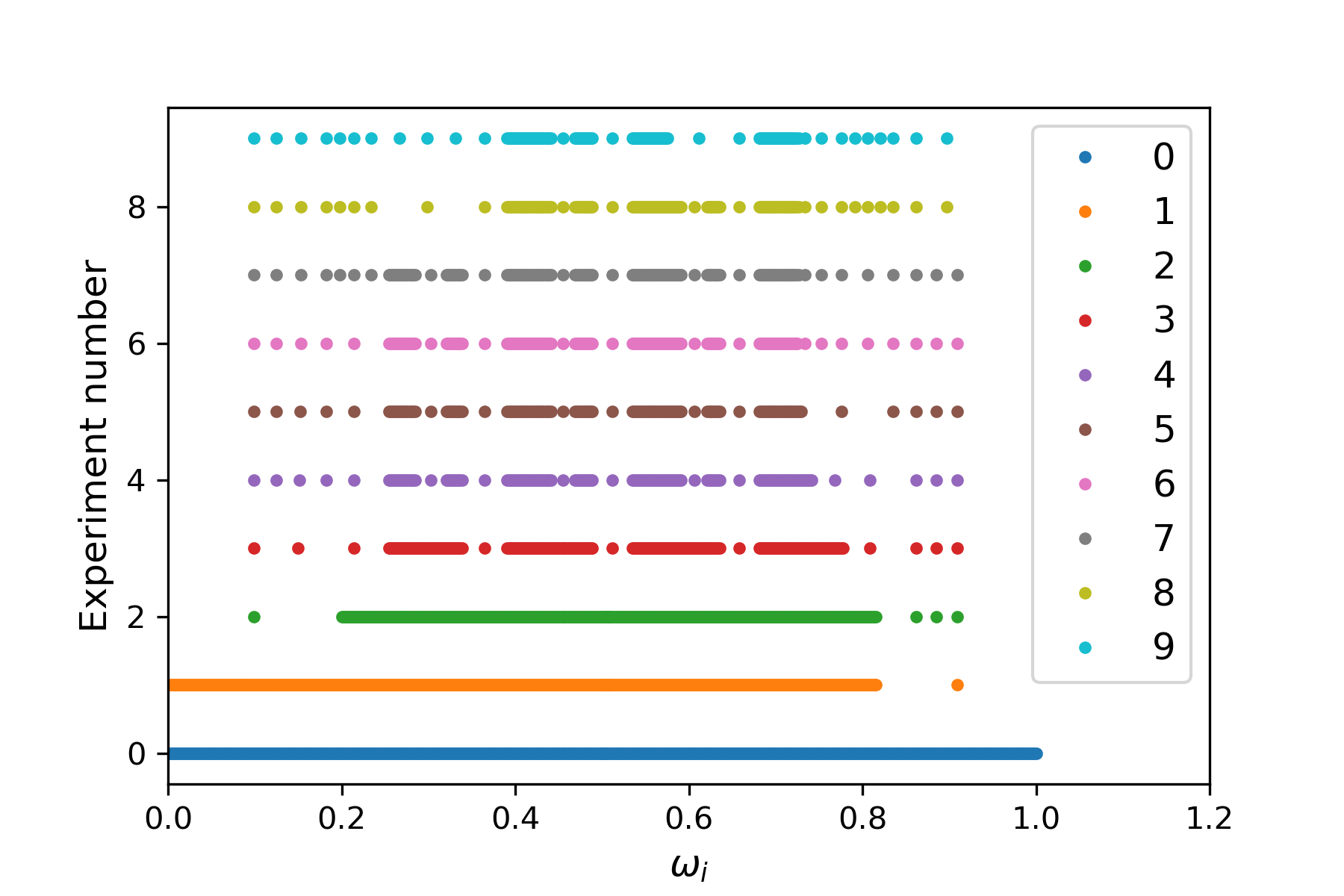}}
\hfill
\subfloat[]{
\includegraphics[width=0.3\linewidth]{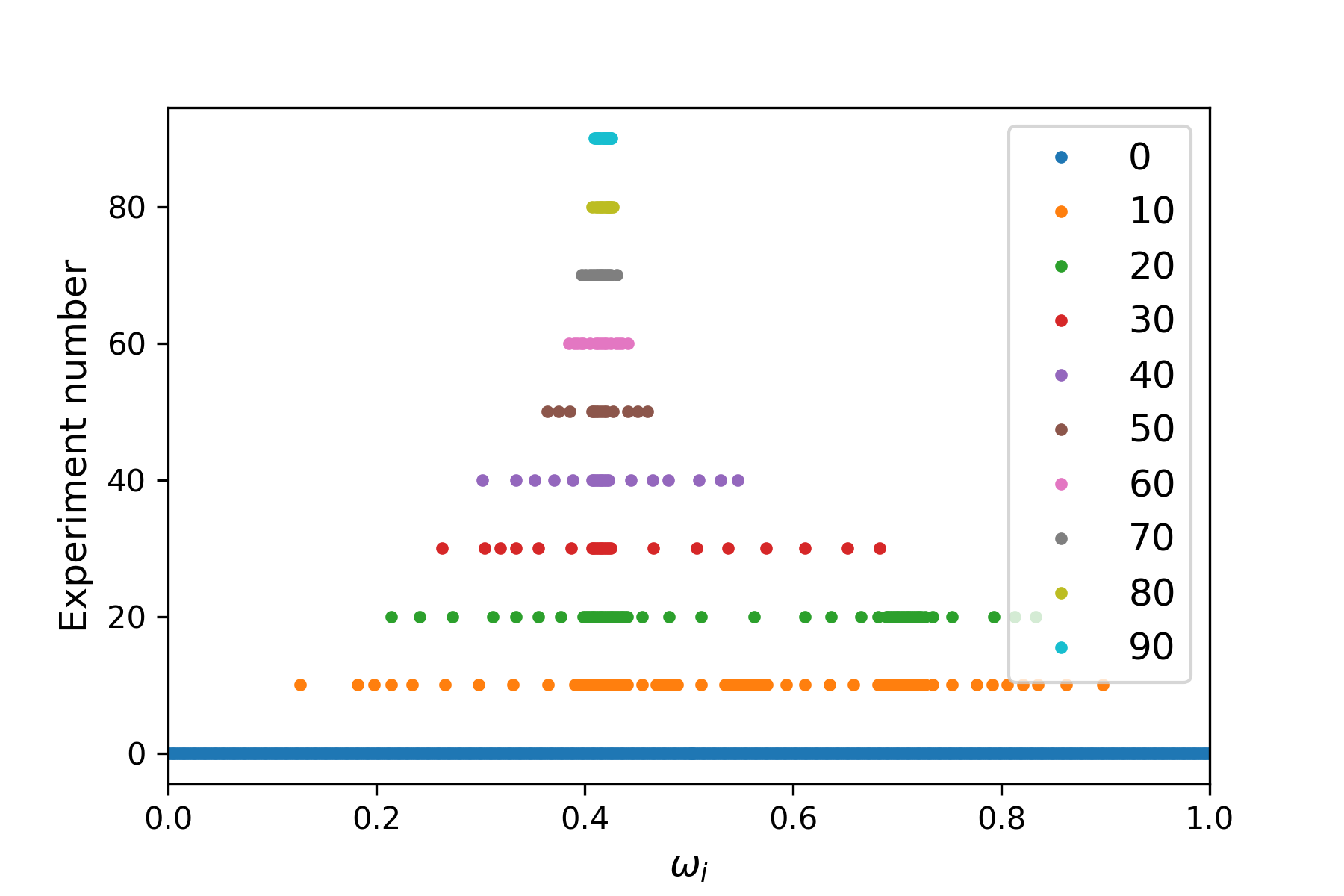}}
\hfill
\subfloat[]{
\includegraphics[width=0.3\linewidth]{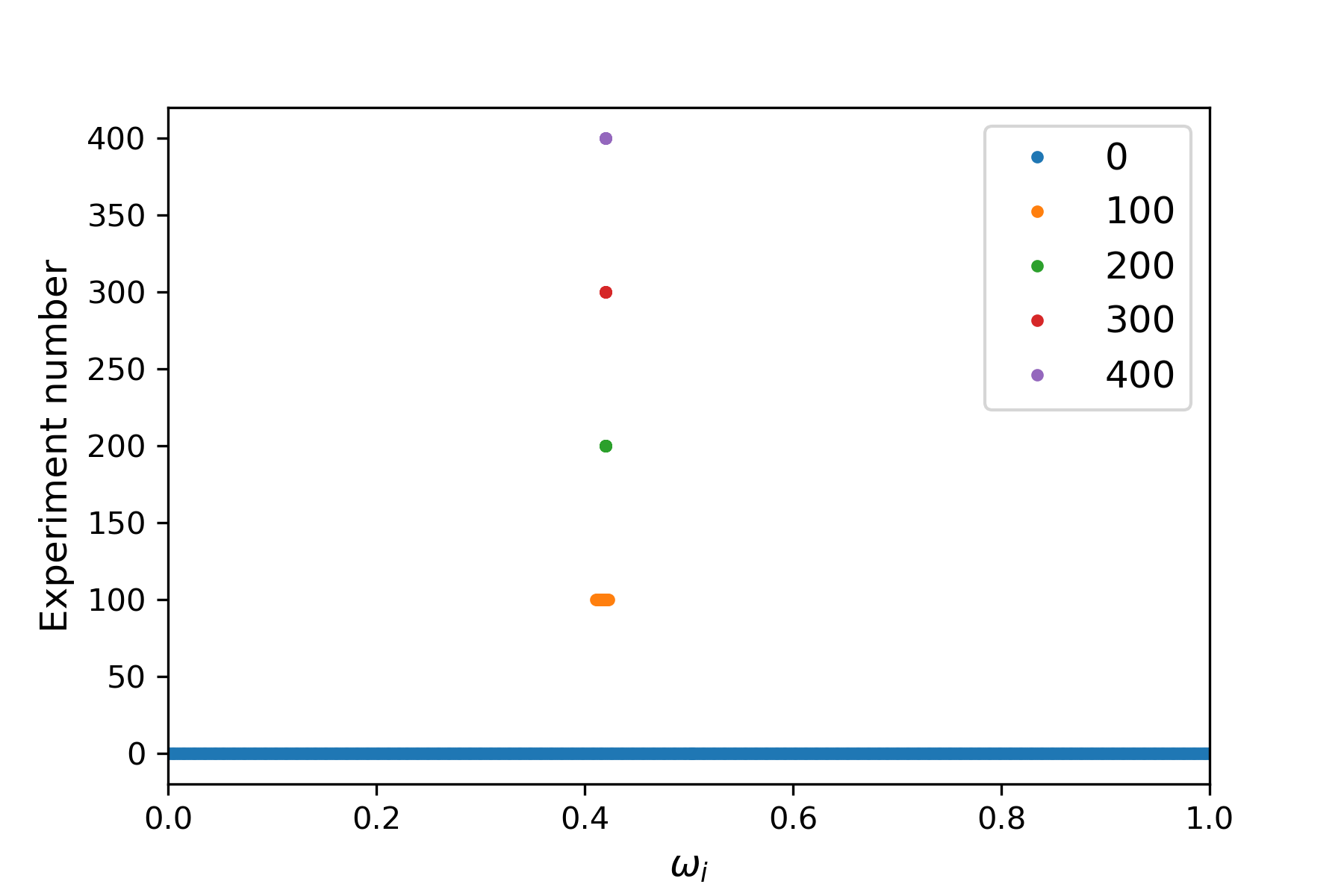}}
\caption{\label{with_1d_grid_refine_steps} Grid refinement ($\omega_i \in [0,1]$) as a function of number of experiments in Bayesian phase estimation $\omega$ in the absence of dephasing noise. $\omega_i$ at experiments a) $0, 1, \cdots, 9$, b) $0, 10, \cdots, 100$ and c) $0, 100, \cdots, 400.$}
\end{figure*}
\begin{figure*}[t!]
\centering
\subfloat[]{
\includegraphics[width=0.3\linewidth]{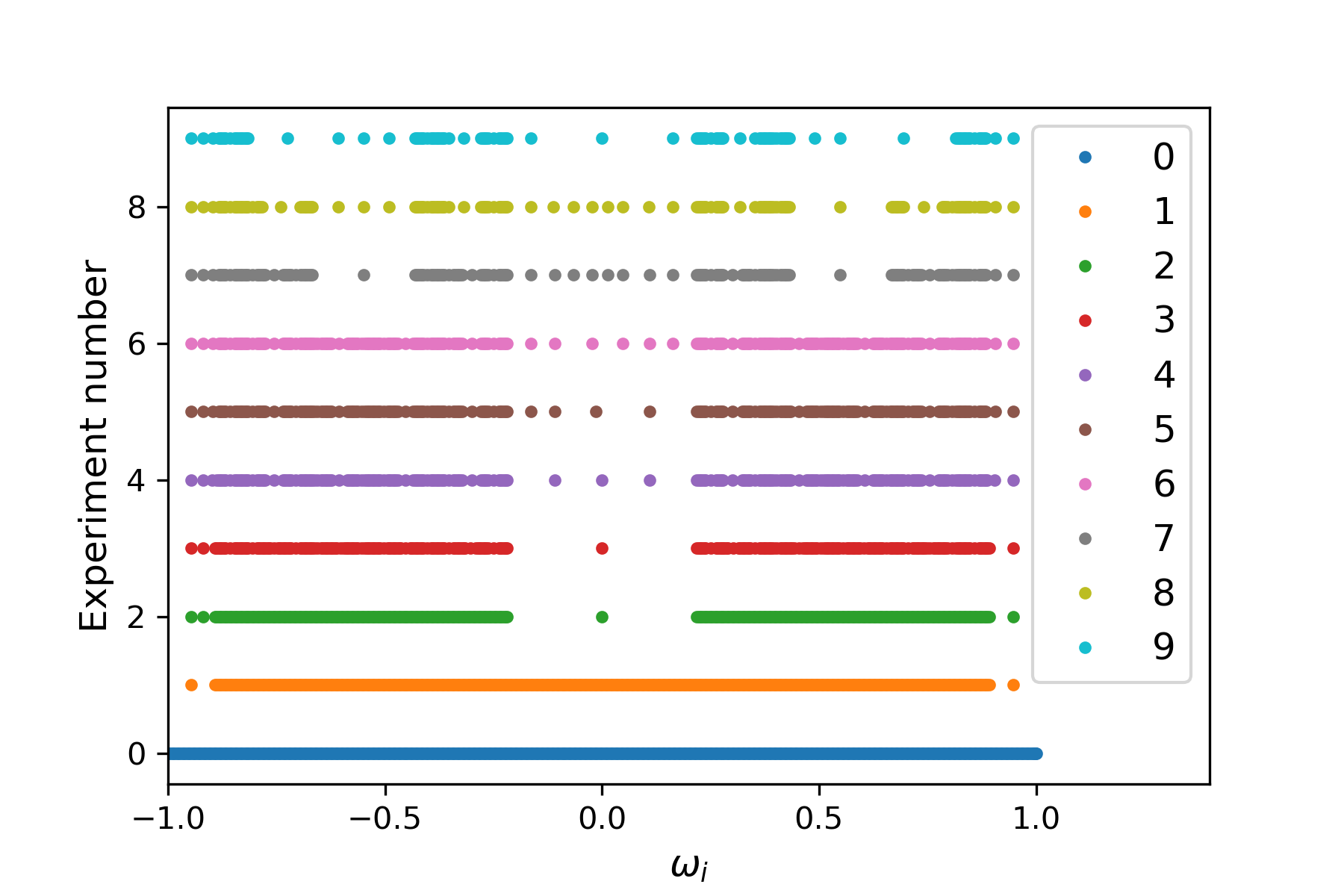}}
\hfill
\subfloat[]{
\includegraphics[width=0.3\linewidth]{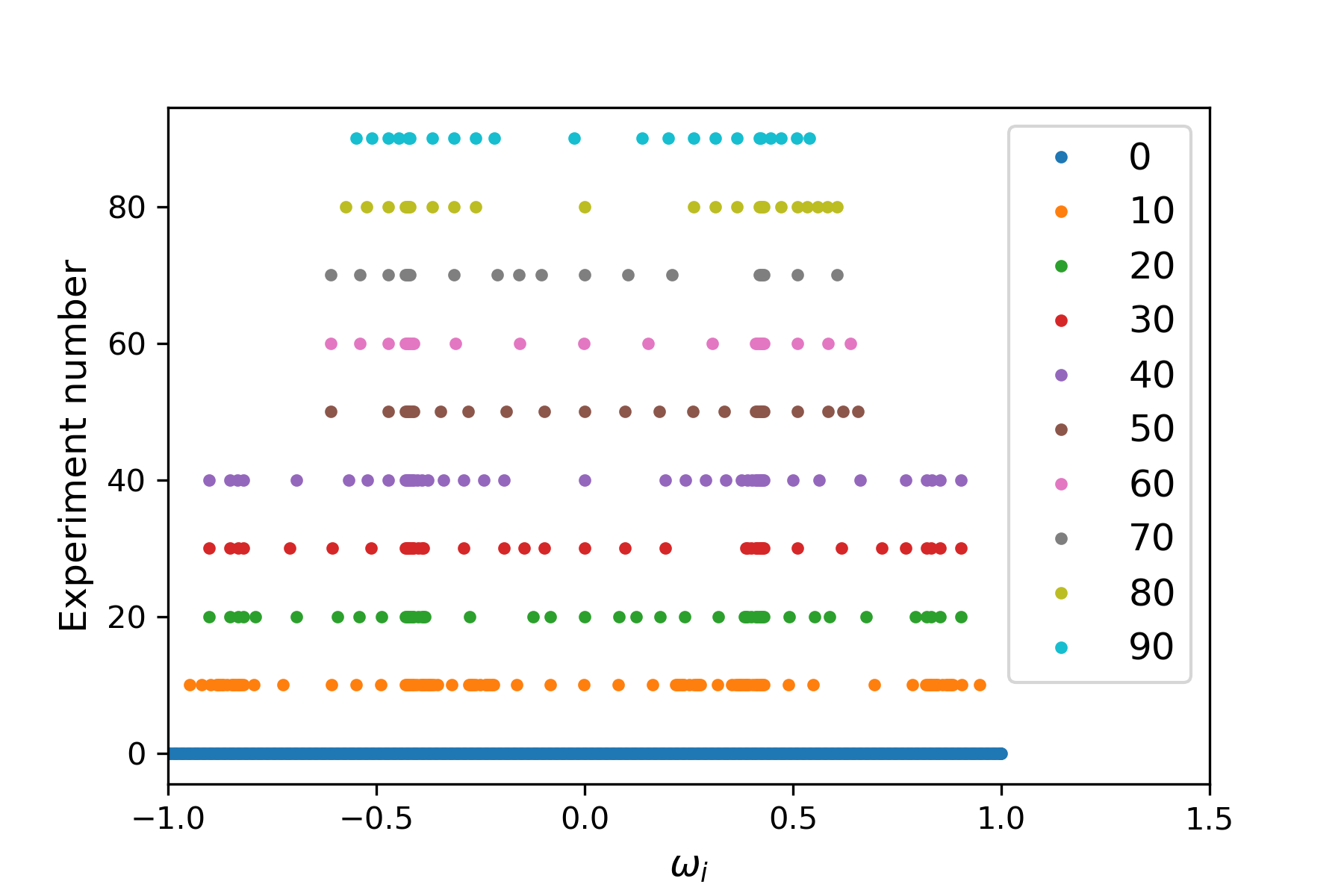}}
\hfill
\subfloat[]{
\includegraphics[width=0.3\linewidth]{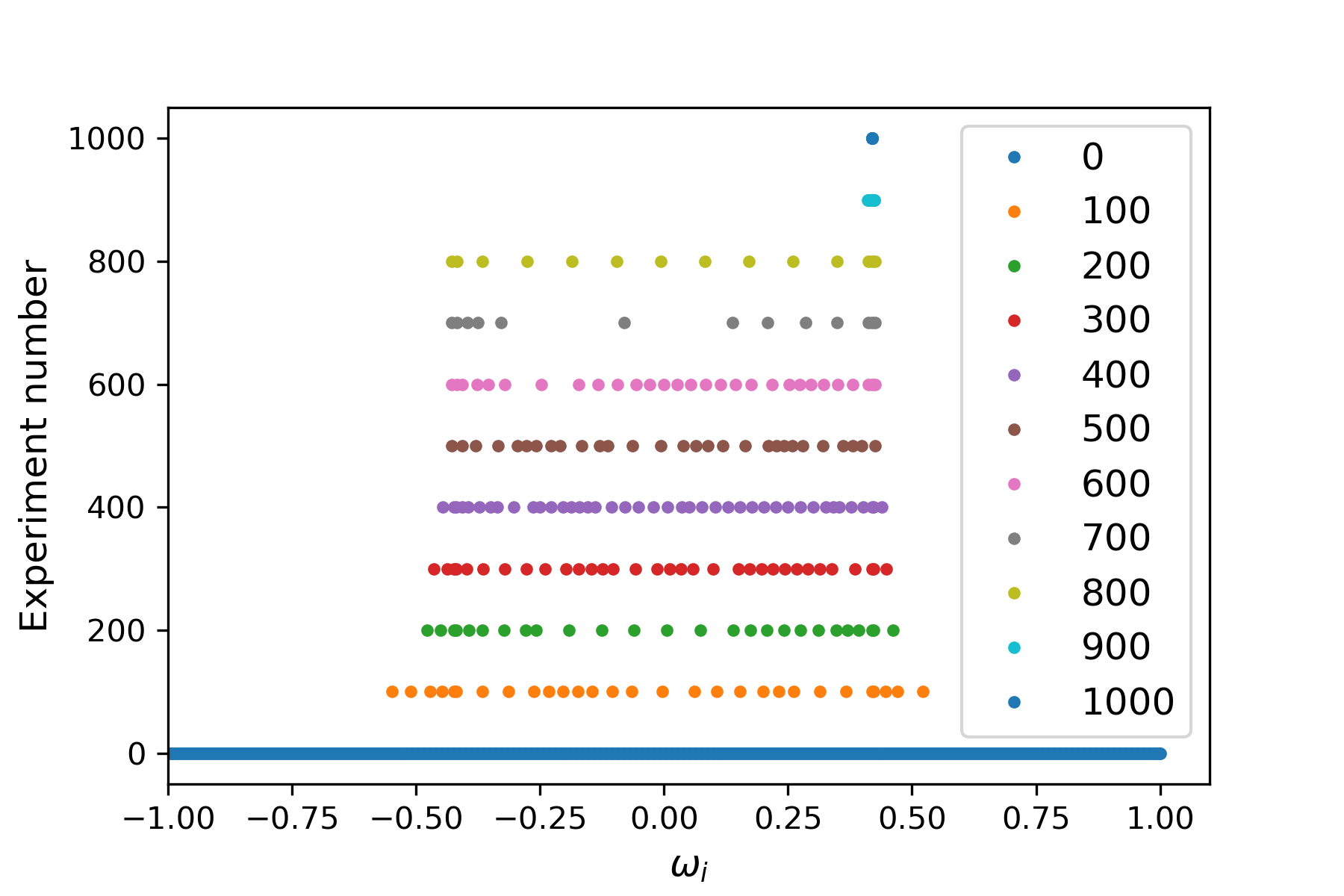}}
\caption{\label{with_1d_grid_refine_omega_neg_steps} Grid refinement ($\omega_i \in [-1,1]$) as a function of number of experiments in Bayesian phase estimation $\omega$ in the absence of dephasing noise. $\omega_i$ at experiments a) $0, 1, \cdots, 9$, b) $0, 10, \cdots, 100$ and c) $0, 100, \cdots, 1000.$}
\end{figure*}
\begin{figure*}[t!]
\centering
\subfloat[]{
\includegraphics[width=0.3\linewidth]{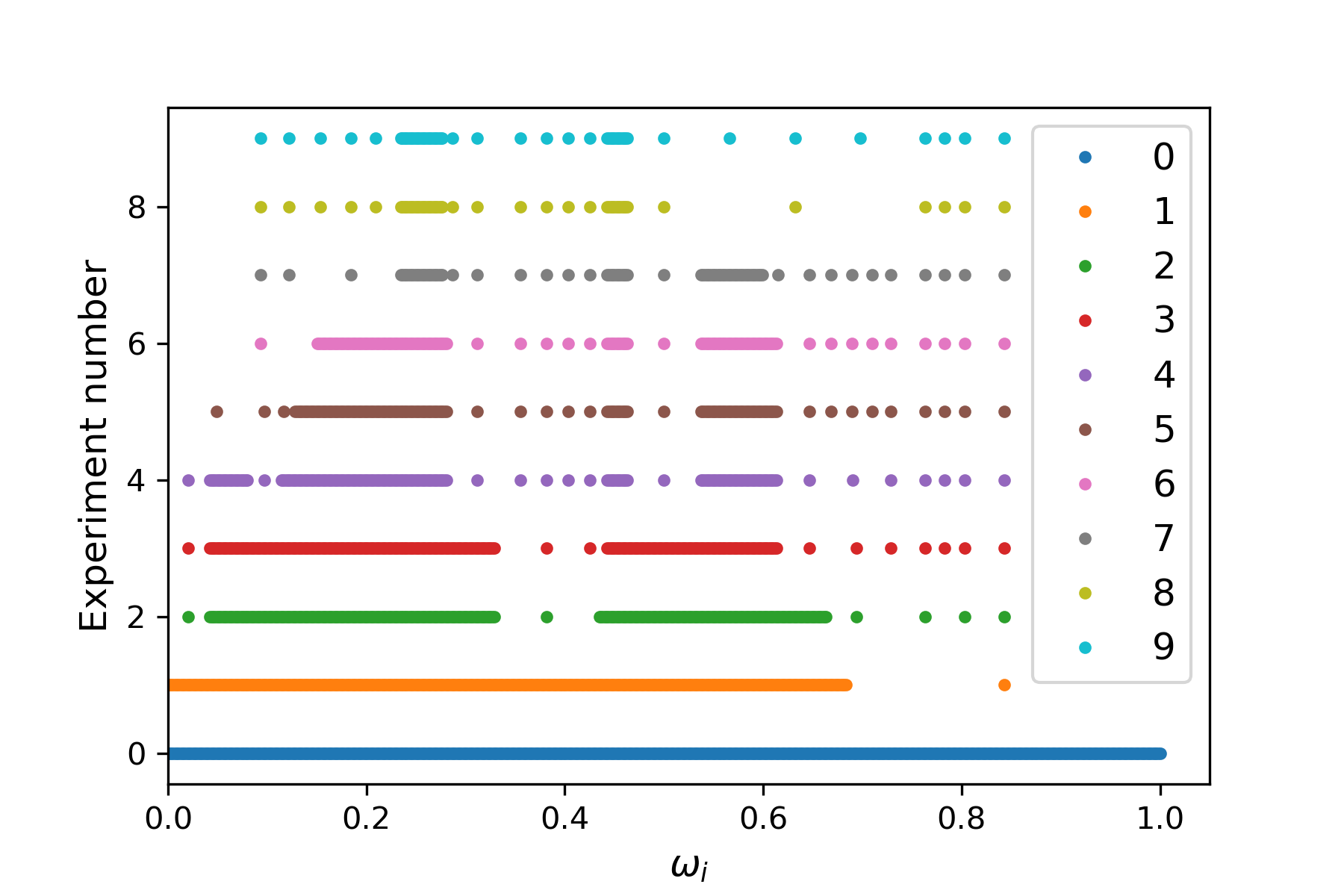}}
\hfill
\subfloat[]{
\includegraphics[width=0.3\linewidth]{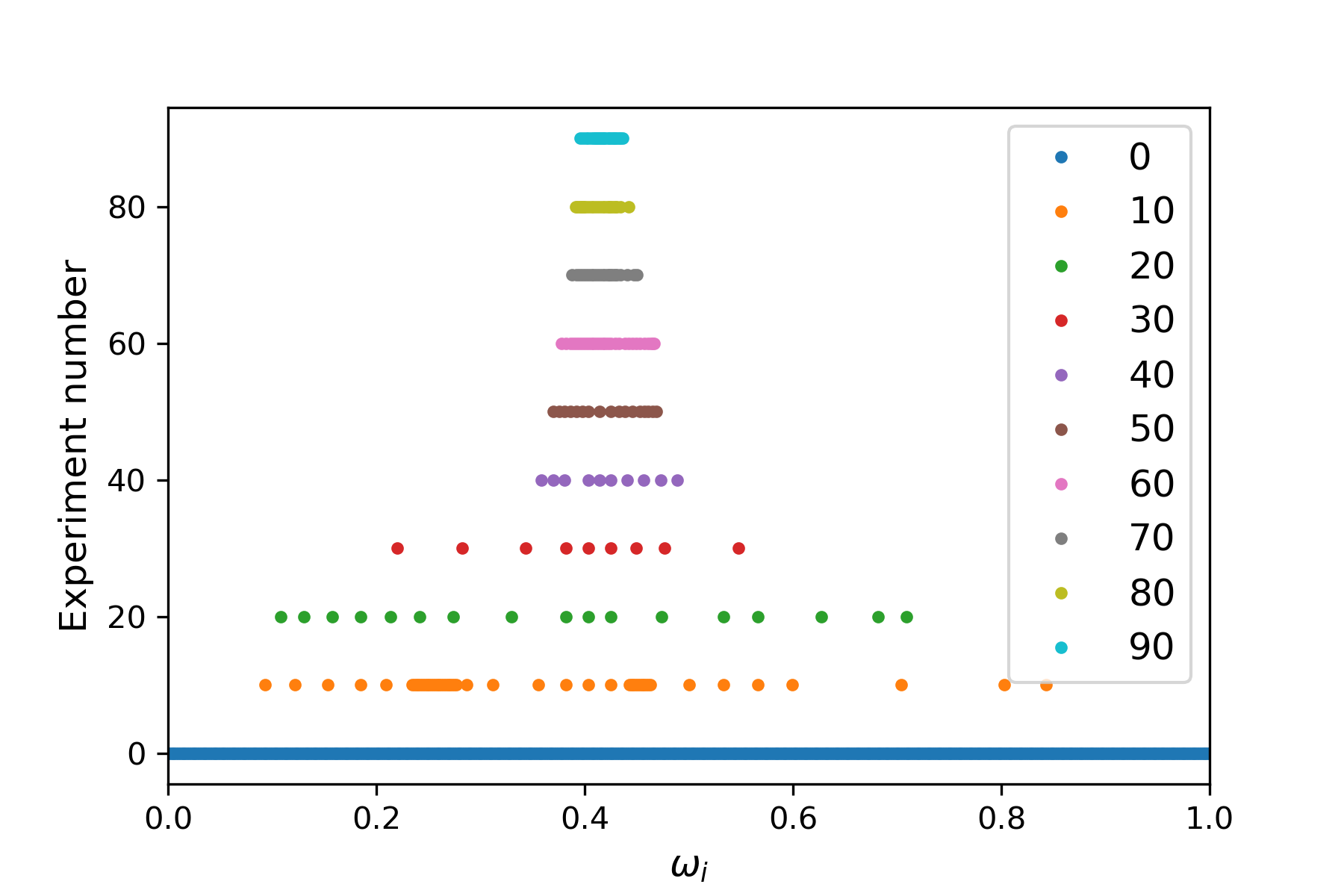}}
\hfill
\subfloat[]{
\includegraphics[width=0.3\linewidth]{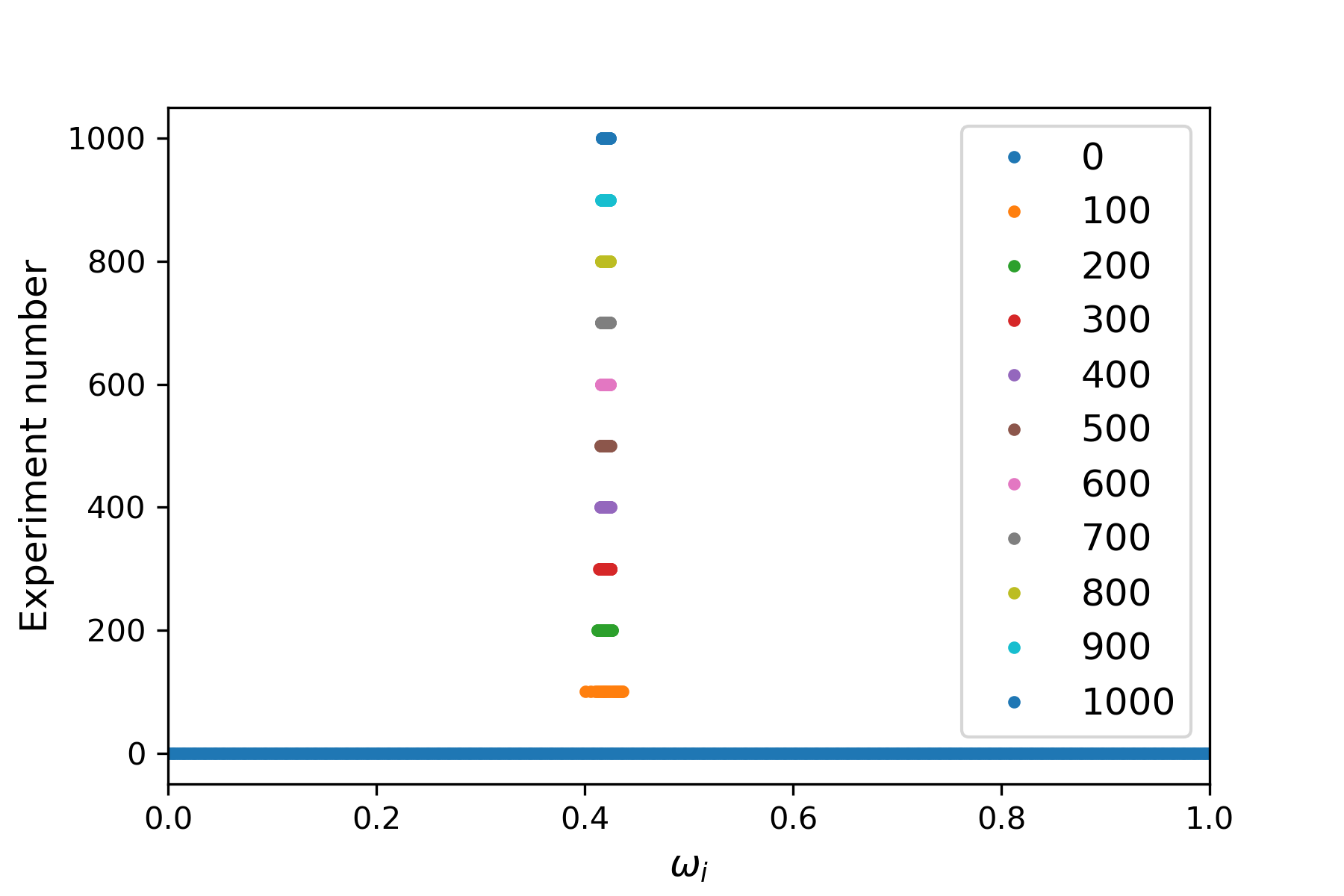}}
\caption{\label{with_dephase_1d_grid_refine_steps} Grid refinement ($\omega_i \in[0,1]$) as a function of number of experiments in Bayesian phase estimation $\omega$ in the presence of dephasing noise with $T_2=100\pi$. $\omega_i$ at experiments a) $0, 1, \cdots, 9$, b) $0, 10, \cdots, 100$ and c) $0, 100, \cdots, 1000.$}
\end{figure*}
In the following we show the progress of adaptive grid refinement phase estimation technique in Figs. \ref{with_1d_grid_refine_steps}, \ref{with_1d_grid_refine_omega_neg_steps} and \ref{with_dephase_1d_grid_refine_steps}. Because of the adaptive grid refinement algorithm~\ref{alg:adapt_refine} and grid merge algorithm~\ref{alg:adaptive_merge} total number particles (grid elements) and the range of the grid changes at each experimental step. Grid refinement algorithm divides the grid elements corresponds to high likelihood and grid merge algorithm merges all he adjacent grid elements with lower likelihood. Repeated application of these two algorithms result in a reduced number of particles concentrated around a small region of very high likelihood. We can see this progression in Figs. \ref{with_1d_grid_refine_steps}, \ref{with_1d_grid_refine_omega_neg_steps} and \ref{with_dephase_1d_grid_refine_steps}. Fig. \ref{with_1d_grid_refine_steps}a shows the grid elements at experiments $0, 1, \dots, 9.$ Although we start with a uniform grid in the region $\omega_i \in [0,1]$, at the end of $9^{th}$ step we see sparsity in some regions concentration in other regions of the domain $[0,1]$. Fig. \ref{with_1d_grid_refine_steps}b shows the remaining grid elements after experiments $0, 10, \cdots, 100.$ We can see concentration of particles around a small region at the end of $100^{th}$ step. Fig. \ref{with_1d_grid_refine_steps}c shows the grid elements after $0, 100, \cdots, 400.$ By the end of $400^{th}$ step, particles are concentrated in a very tiny region within the domain $[0,1]$. These figures show how the adaptive grid refinement and merge algorithms combined with Bayesian method can be used effectively for quantum phase estimation. Figs. \ref{with_1d_grid_refine_omega_neg_steps}a, \ref{with_1d_grid_refine_omega_neg_steps}b and \ref{with_1d_grid_refine_omega_neg_steps}c show the grid refinement after the experiments $0, 1, \dots, 9,$ $0, 10, \cdots, 100.$ and $0, 100, \cdots, 1000.$ steps respectively for quantum phase estimation for $\omega \in [-1,1]$ without dephasing noise. Unlike the previous case when $\omega \in [0,1],$ the range of the particles narrows down at around $900$ experiments. It shows that as the complexity increases, number of experiments needed and hence the computational cost increases. The complexity of the problem further increases with the introduction of the dephasing noise.
Figs. \ref{with_dephase_1d_grid_refine_steps}a, \ref{with_dephase_1d_grid_refine_steps}b and \ref{with_dephase_1d_grid_refine_steps}c show the grid refinement after the experiments $0, 1, \dots, 9,$ $0, 10, \cdots, 100.$ and $0, 100, \cdots, 1000.$ steps respectively for quantum phase estimation in the presence of dephasing noise with $T=100\pi.$ As with any other phase estimation technique dephasing noise deteriorates the accuracy of the estimated phase. Hence we continued for a larger number ($1000$) of  experiments instead of $400.$ Although we need higher number of experiments as we increase the complexity of the problem, the adaptive grid refinement Bayesian phase estimation converges to the correct estimate with desired accuracy. However the LW resampling algorithm fails for certain cases and the correct estimate can never be achieved even when we increases the number of experiments or the number of particles.


\end{document}